\documentclass[final]{IEEEtran}
 \pdfoutput=1
\usepackage{amssymb}
\usepackage{multirow}
\usepackage{graphicx}
\usepackage[cmex10]{amsmath}
\usepackage{subfigure}
\usepackage{times}
\usepackage{comment}
\usepackage{multirow}
\usepackage{ifpdf}
\usepackage{amsfonts}
\usepackage{cite}
\usepackage{listings}
\usepackage{xcolor}
\usepackage{url}

\begin{document}
%
\title{Rate Model for Compressed Video Considering Impacts Of Spatial, Temporal and Amplitude Resolutions and Its Applications for \\Video Coding and Adaptation}

%
%

\author{Zhan Ma, Hao Hu, Meng Xu, and Yao Wang\\
\thanks{Z. Ma was with the Polytechnic Institute of New York University, Brooklyn, NY 11201, USA.
He is now with the Dallas Technology Lab, Samsung Telecommunications America, Richardson,
TX 75082, USA. (email: zhan.ma@ieee.org)}
\thanks{H. Hu, M. Xu and Y. Wang are with the Polytechnic Institute of New York University, Brooklyn, NY 11201, USA.
(email: \{hhu01, mxu02\}@students.poly.edu, yao@poly.edu)}
}


\maketitle

\begin{abstract}
In this paper, we investigate the impacts of spatial, temporal and
amplitude resolution (STAR) on the bit rate of a compressed video.
We propose an analytical rate model in terms of the quantization
stepsize, frame size and frame rate. Experimental results reveal
that the increase of the video rate as the individual resolution
increases follows a power function. Hence, the proposed model
expresses the rate  as the product of power functions of the
quantization stepsize, frame size and frame rate, respectively. The
proposed rate model is analytically tractable, requiring only four
content dependent parameters. We also propose methods for predicting
the model parameters from content features that can be computed from
original video. Simulation results show that model predicted rates
fit the measured data very well with high Pearson correlation (PC)
and small relative root mean square error (RRMSE). The same model
function works for different coding scenarios (including scalable
and non-scalable video, temporal prediction using either
hierarchical B or IPPP structure, etc.) with very high accuracy
(average PC $>$ 0.99), but the values of model parameters differ.
Using the proposed rate model  and the quality model introduced in a
separate work, we show how to optimize the STAR for a given rate
constraint, which is important for both encoder rate control and
scalable video adaptation. Furthermore, we demonstrate how to order
the spatial, temporal and amplitude  layers of a scalable video in a
rate-quality optimized way.
\end{abstract}



%
\section{Introduction} \label{sec:intro}
 A fundamental and challenging problem in video encoding is, given a
target bit rate, how to determine at which spatial
resolution (i.e., frame size or FS), temporal resolution (i.e.,  frame rate or FR), and
amplitude resolution (usually controlled by the
quantization stepsize or QS), to code the video. One may code the
video at a high FR, large FS, but high QS,
yielding noticeable coding artifacts in each coded frame.
Or one may use a low FR, small FS, but small QS,
producing high quality frames. These and other
combinations can lead to very different perceptual quality.
Ideally, the encoder should choose the spatial, temporal, and amplitude resolution (STAR) that leads to the best perceptual
quality, while meeting the target bit rate. Optimal solution requires
accurate rate and perceptual quality prediction at any STAR combination.

In this paper, 
we investigate how does the rate change as a function of the quantization stepsize $q$, frame rate $t$
and frame size $s$. This work
is extended from our previous work~\cite{Ma_RateQualityModel},
where we consider the impact of temporal and amplitude resolutions on the video rate.
Rate modeling for video coding has been researched over decades.
To
the best of our knowledge, no prior work has considered the joint
impact of frame size, frame rate and quantization on the bit rate. However,
several prior works~\cite{RC_RQModel,YQZHANG_RC,RibasLei_RC,He_LinearModel}
have considered rate modeling under a fixed frame rate and fixed frame size,
and have proposed models that relate the average bit rate
versus  quantization stepsize $q$. Please see~\cite{Ma_RateQualityModel}
for a review of prior work on this subject.


The proposed model is derived based on our analysis of the actual rates of video when coded at
different ($q$, $s$, $t$) combinations. Our analysis shows that the rate at  any ($q$, $s$, $t$) can
be approximated well by the product of three separable functions, representing the influence
of $q$, $s$, $t$ on the rate, respectively. Each function can be approximated well by a power
function, and has a single content-dependent model parameter.  The overall model has four parameters (including
the maximum bit rate $R_{\max}$) and fits the measured rates for
different STAR very
accurately (with an average PC larger than 0.99 over all
sequences). We also investigate how to predict the parameters
using content features. According to our experiments,
the model parameters can be
estimated very well  from three content features only, with average PC larger than
0.99.

The main contributions of this paper include:
\begin{itemize}
\item This is the first work modeling the compressed video bit rate
      with respect to the video FS $s$, FR $t$ and QS $q$. The proposed model is analytically simple and requires only four content
      dependent model parameters.
\item We also develop efficient method for predicting model parameters using a few content features extracted
      from the raw video. Our proposed model does not require any off-line training
      process.
\item The proposed  rate model, together with the quality model presented elsewhere~\cite{QSTAR},
      make rate-quality optimized scalable video
      adaptation and encoder rate control problems analytical tractable. We also
      develop a quality optimized layer ordering algorithm, which facilitates simple scalable video rate adaptation
      at a network proxy or gateway to maximize the streamed video quality
      given the bit rate constraint.
\end{itemize}

The remainder of this paper is organized as follows:
Section~\ref{sec:r_star} presents the rate model considering the
joint impact of spatial, temporal and amplitude resolutions. We then
validate the  same rate model is applicable for different encoding
settings in Section~\ref{sec:model_validate}. Model parameter
prediction using content features is analyzed in
Section~\ref{sec:para_pred}, while Section~\ref{sec:svc_adapt}
introduces the rate-constrained quality and STAR optimization for
both encoder rate control and scalable video adaptation, and
considers layer ordering for scalable video in a rate-quality
optimized way. Section~\ref{sec:conclusion} concludes the current
work and discusses the future research directions.

\section{Rate Model Considering Impacts of Spatial, Temporal and Amplitude Resolutions}
\label{sec:r_star} In this section, we develop the rate
model based on the rates of video bitstreams generated using the
spatial and temporal scalability of SVC at multiple fixed
quantization parameters (QPs).

\subsection{Normalized Rate Data Collection and Analysis}

To see how QS, FS, and FR individually affect the bit rate, we
define the following normalized rate functions:
$R_s(s; q, t) = {R(q, s, t)}/{R(q, s_{\max}, t)}$
is normalized rate versus spatial resolution (NRS) under a certain $q$ and $t$;
$R_q(q; s, t) = {R(q, s, t)}/{R(q_{\min}, s, t)}$
is the normalized rate versus quantization stepsize (NRQ) under a
certain $s$ and $t$; and
 $R_t(t; q, s)={R(q, s, t)}/{R(q,s, t_{\max})}$
is the normalized rate versus temporal resolution (NRT) under given
$q$ and $s$.

Note that  the NRT function $R_t(t; q, s)$ describes how does the
rate decrease when the frame rate $t$ reduces from $t_{\max}$ while
the FS and QS are fixed; NRQ function $R_q(q; s, t)$ illustrates how
does the rate change when the quantization stepsize increases beyond
$q_{\min}$ under constant $s$ and $t$; while the NRS function
$R_s(s; q, t)$ characterizes how does the rate reduce when the frame
size decreases from $s_{\max}$ when $q$ and $t$ are fixed.

\begin{figure*}[htp]
    \centering
\includegraphics[scale=0.54]{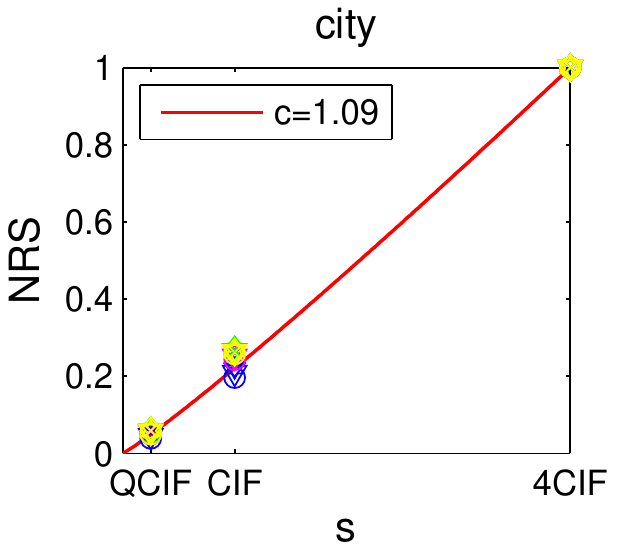}
\includegraphics[scale=0.54]{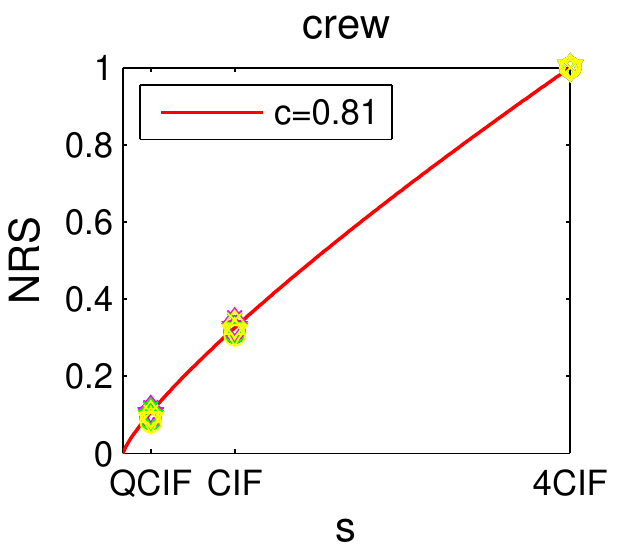}
\includegraphics[scale=0.54]{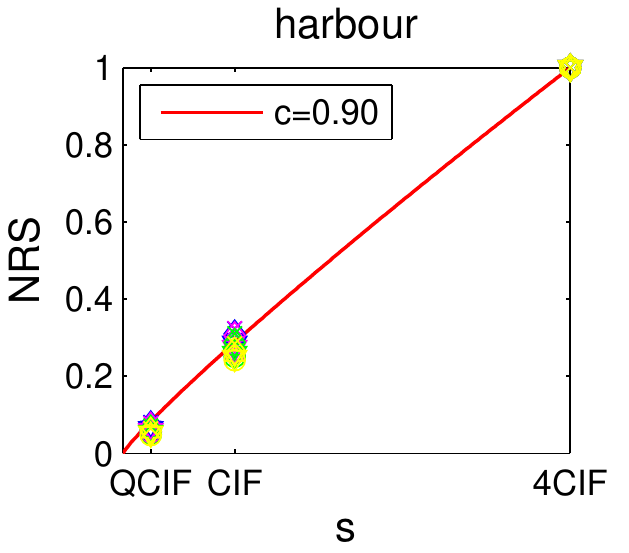}
\includegraphics[scale=0.54]{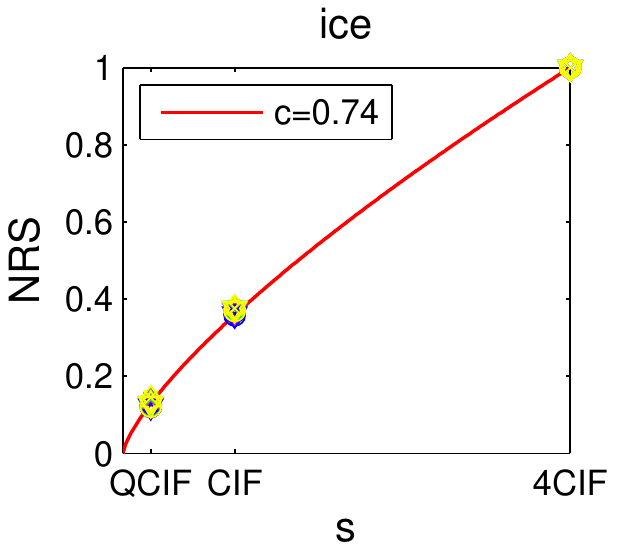}
\includegraphics[scale=0.54]{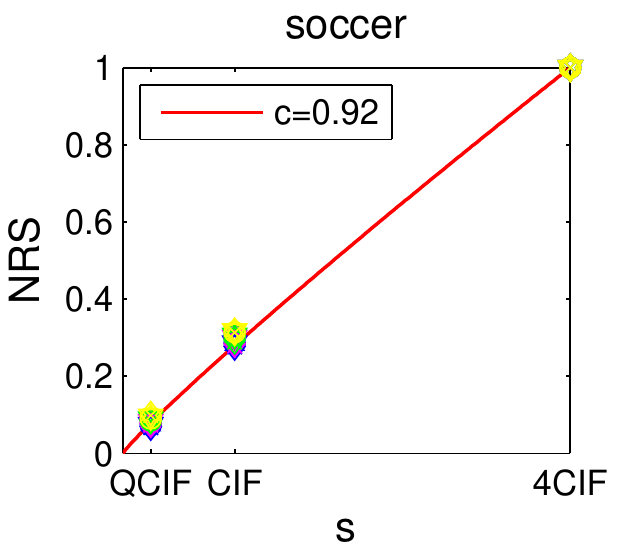}

\includegraphics[scale=0.54]{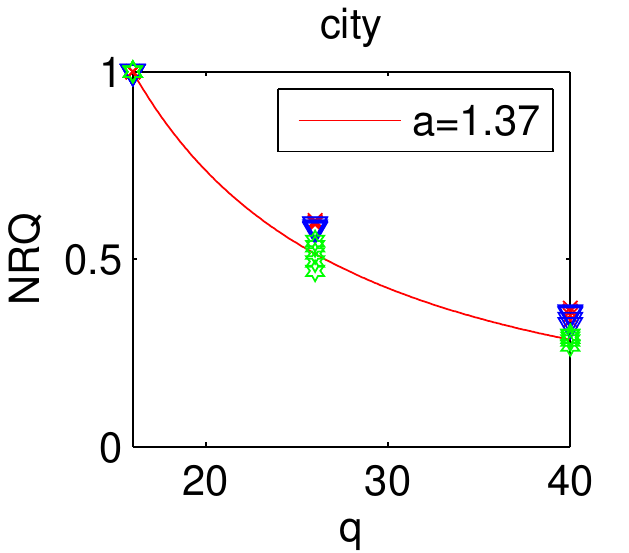}
\includegraphics[scale=0.54]{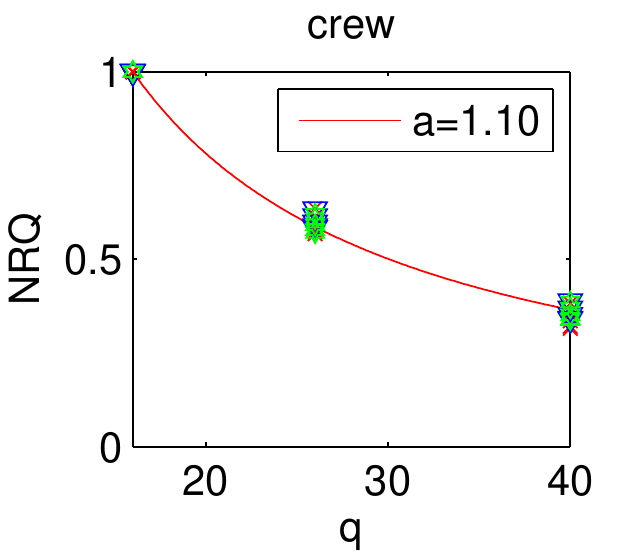}
\includegraphics[scale=0.54]{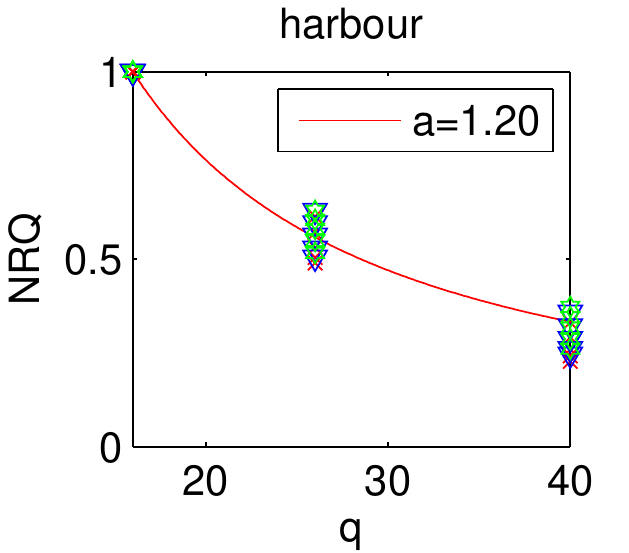}
\includegraphics[scale=0.54]{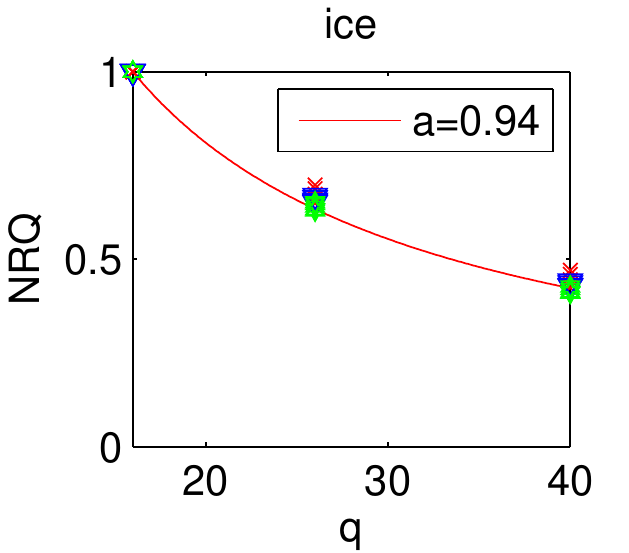}
\includegraphics[scale=0.54]{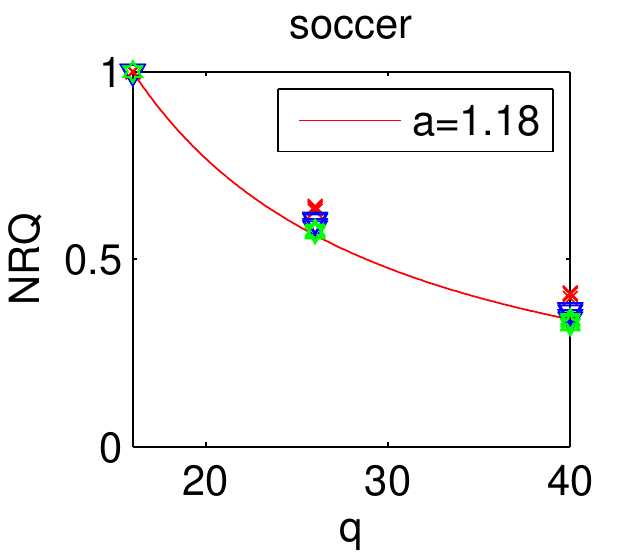}

\includegraphics[scale=0.54]{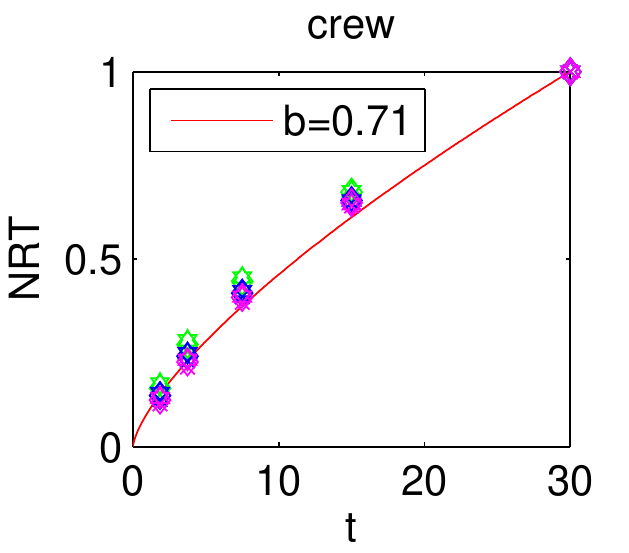}
\includegraphics[scale=0.54]{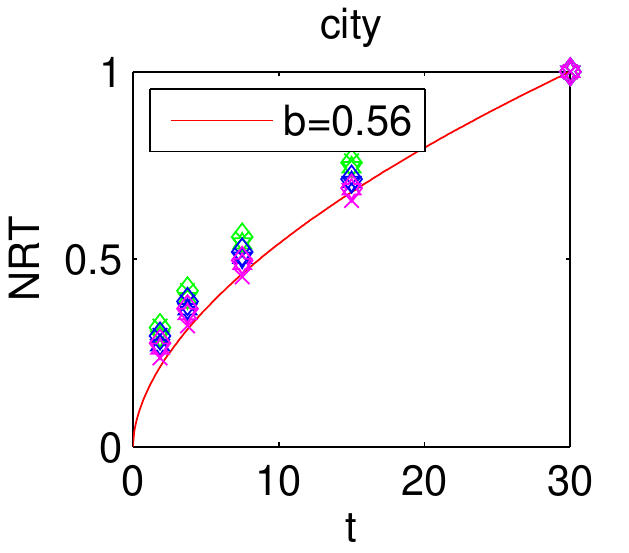}
\includegraphics[scale=0.54]{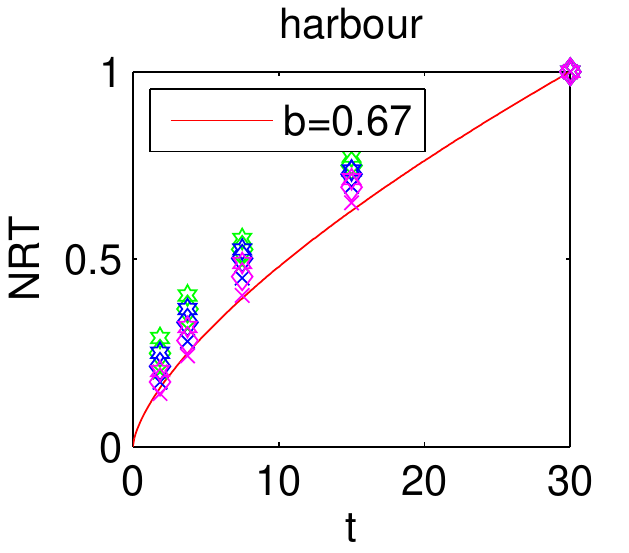}
\includegraphics[scale=0.54]{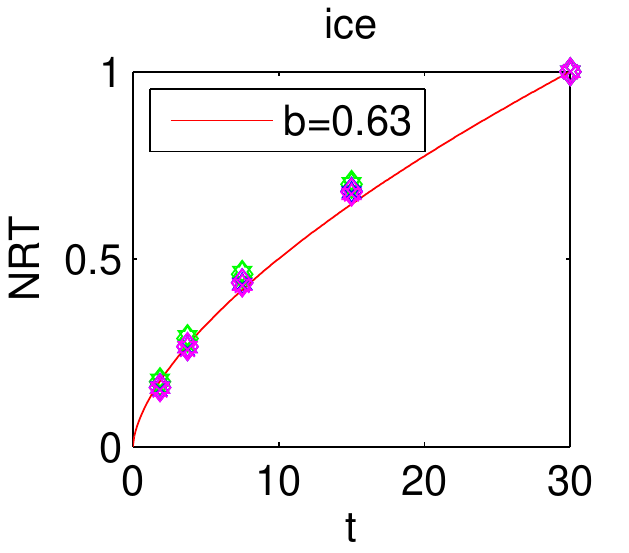}
\includegraphics[scale=0.54]{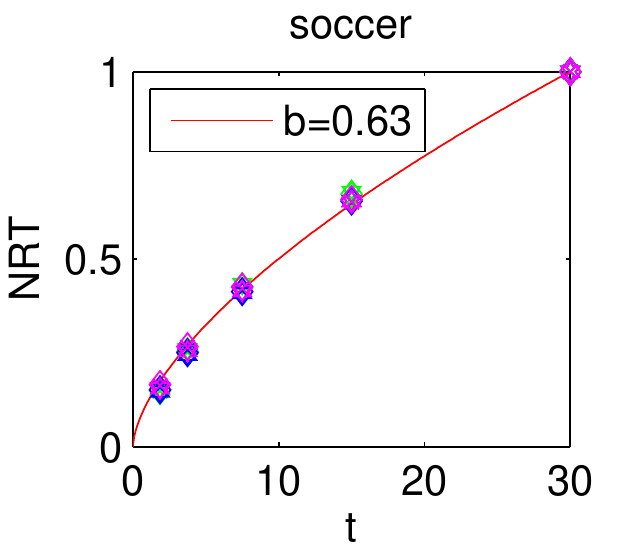}

    \caption{Illustrations of NRS, NRQ and NRT for
    all combinations of $q$, $s$ and $t$, where $q\in[64, 40, 26, 16]$,
    $t\in[1.875, 3.75, 7.5, 15, 30]$ and $s\in~$[QCIF, CIF, 4CIF].
    Points are measured rates, curves are predicted rates using respective
    Eqs.~(\ref{eq:SCFR})~(\ref{eq:NRQ}) and~(\ref{eq:NRT}). NRS curves
    are fitted using all possible $q$ and $t$; NRQ curves are fitted
    using all possible $t$ at $s_{\max}$ (green hexagram markers), while
    NRT curves are obtained using points at $q_{\min}$ and $s_{\max}$ (magenta cross markers).
    In our case, $q_{\min}$ = 16, $t_{\max}$ = 30 Hz and $s_{\max}$ = 4CIF.
    }
\label{fig:NRS}
\end{figure*}

To see how NRS, NRQ, and NRT vary with $s$, $q$, and $t$,
respectively, we encoded several test videos using  the joint
spatial and temporal scalability tool of
 the SVC reference software JSVM~\cite{JSVM} and
 measured the actual bit rates corresponding to different STARs. Specifically, five
 video sequences,  ``city'', ``crew'', ``harbour'', ``ice'' and ``soccer'', at
 original 4CIF (704x576) resolution, are encoded
into 5 temporal layers using dyadic hierarchical prediction
structure, with frame rates at 1.875, 3.75, 7.5, 15 and 30 Hz,
respectively, and each temporal layer contains 3 dyadic spatial
layers (i.e., QCIF, CIF and 4CIF). For simplicity, we  apply the
same QP for all temporal and spatial layers (i.e., without using
{\it QP cascading}~\cite{JSVMEncoderDescp}). To investigate the
impact of QP, we have coded the video using QP ranging from 16 to
44. Here, we only present the results with QP = 40, 36, 32 and 28.
Corresponding quantization stepsizes are 64, 40, 26 and 16,
respectively~\cite{H264std}. Other QPs have the similar performance
according to our simulation results.

The bit rates of all layers are collected and normalized by the rate
at the largest frame size, i.e., 4CIF , to find NRS points $R_s(s;
q, t)$ for all $q$ and $t$, which are plotted in the first row of
Fig.~\ref{fig:NRS}. As shown, the NRS curves obtained with different
$q$ and $t$ overlap with each other, and can be captured by a single
curve quite well. Similarly, the NRQ curves  (middle row) are also
almost invariant with the frame rate $t$, and vary slightly for
different frame size $s$ as shown in Fig.~\ref{fig:NRS}; On the
other hand, NRT curves (last row) are quite dependent on frame size
and quantization as shown in Fig.~\ref{fig:NRS}.

To derive the overall rate model, we recognize that the rate
function $R(q,s,t)$ can be decomposed  in multiple ways as follows:
\begin{subequations}
\begin{align}
  &R(q,s,t) \nonumber\\
  &=R_{\max}R_q(q;s_{\max},t_{\max})R_s(s; q,t_{\max})R_t(t;q,s),\label{eq:rstar_decomp1} \\
          &=R_{\max}R_q(q;s_{\max},t_{\max})R_t(t;q,s_{\max})R_s(s; q,t), \\
          &=R_{\max}R_s(s;q_{\min},t_{\max})R_t(t;q,s_{\max})R_s(s;q,t), \\
          &=R_{\max}R_s(s;q_{\min},t_{\max})R_s(s;q,t_{\max})R_t(t;q,s), \\
          &=R_{\max}R_t(t;q_{\min},s_{\max})R_s(s;q_{\min},t)R_q(q;s,t), \\
          &=R_{\max}R_t(t;q_{\min},s_{\max})R_q(q; s_{\max},t)R_s(s;q,t). \label{eq:rstar_decomp}
          \end{align}
\end{subequations}

Among all these feasible decompositions, we should choose the one
that leads to simplest mathematical model.   For example, if we
choose the first decomposition in (\ref{eq:rstar_decomp1}), the
dependency of $R_t(t;s,q)$ on $s$ and  $q$ makes the overall model
complicated. On the other hand, if we choose the one in
(\ref{eq:rstar_decomp}), the fact that $R_s(s;q,t)$ does not depend
on $q$ and $t$ , and $R_q(q; s_{\max}, t)$ does not depend on $t$ ,
enables us to write (\ref{eq:rstar_decomp}) as the product of three
separable functions of $t$, $q$, $s$, respectively, i.e.,
\begin{align}
R(q,s,t)=R_{\max} \tilde{R}_t(t) \tilde{R}_q(q) \tilde{R}_s(s),
\label{eq:rstar_abstraction}
\end{align}
 where $\tilde{R}_t(t)$ denotes a model for the $R_t(t;q_{\min},s_{\max})$ data that
depends on $t$ only;  $\tilde{R}_q(q)$  indicates a model for
$R_q(q;s_{\max},t)$ data that depends on $q$ only; and finally
$\tilde{R}_s(s)$ represents a model for the $R_s(s;q,t)$ data,
depending on $s$ only. With the general form in
(\ref{eq:rstar_abstraction}) as  the proposed rate model, the
remaining rate modeling problem is divided into three parts. One is
to devise an appropriate functional form for $\tilde{R}_{s}(s)$, so
that it can model the measured NRS points for all $q$ and $t$ in
Fig.~\ref{fig:NRS} accurately; the second one is to derive an
appropriate functional form for $\tilde{R}_q(q)$ that can accurately
model the measured NRQ points for all $t$ at $s=s_{\max}$, and the
third part is to provide a proper functional form for
$\tilde{R}_t(t)$ that can accurately capture the measured NRT points
at $q_{\min}$ and $s_{\max}$.

\subsection{Model for Normalized Rate v.s. Quantization $\tilde{R}_q(q)$}

$\tilde{R}_q(q)$ is used to describe the reduction of the normalized
bit rate as the QS increases at a fixed frame size $s_{\max}$ for
any given frame rate $t$. As shown in Fig.~\ref{fig:NRS} (middle row),
$\tilde{R}_q(q)$ is independent of the $t$, thus we can model the
$\tilde{R}_q(q)$ at any frame rate (e.g., $t$ = 30 Hz) for
simplicity. This reduces the problem to model the influence of QS
on the bit rate under fixed FS and FR, which has been studied
extensively. For example,  Altunbasak {\it et al}
\cite{Altunbasak_DCT} demonstrate that the rate can be approximated
by  a power function of $q$  if the transform coefficients to be
quantized follow the Cauchy distribution. On the other hand, Ding
and Liu~\cite{RC_RQModel} show the power function is also a good
rate estimation model for Gaussian distributed source.

Following these prior work, we also assume that $R(q)$ follows a
power function with the general form of
\begin{align}
  R(q) = \frac{\theta}{q^a},
\end{align} where $\theta$ and $a$ is content related parameters.  Usually,
$\theta$ is related to the picture content complexity. Assuming the
same $\theta$ for different quantization, normalized rate versus
quantization can be approximated by
\begin{align}
  \tilde{R}_q(q) = \frac{R(q)}{R(q_{\min})}= \left(\frac{q}{q_{\min}}\right)^{-a}. \label{eq:NRQ}
\end{align} Fig.~\ref{fig:NRS} (middle row) shows that
that the model (\ref{eq:NRQ}) fits the measured data points
accurately. The parameter $a$ is determined by minimizing the
squared error, and it characterizes how
fast the bit rate reduces when $q$ increases. 
We note that
the model in (\ref{eq:NRQ}) is consistent with  the
model proposed by Ding and Liu~\cite{RC_RQModel} for non-scalable
video at fixed frame size and frame rate, where they have found that the parameter $a$ is in the range
of 0-2.

\subsection{Model for Normalized Rate v.s. Temporal Resolution $\tilde{R}_t(t)$}
As explained earlier, $\tilde{R}_t(t)$ is used to describe the
reduction of the normalized bit rate as the frame rate reduces at
$q_{\min}$ and $s_{\max}$. Therefore,  the desired property for the
$\tilde{R}_t(t)$ function is that it should be 1 at $t = t_{\max}$
and monotonically reduces to 0 at $t = 0$. Based on the data trend
shown in Fig.~\ref{fig:NRS} (third row), we choose a power function
to describe the $\tilde{R}_t(t)$, i.e.,
\begin{align}
  \tilde{R}_t(t) = \left(\frac{t}{t_{\max}}\right)^b. \label{eq:NRT}
\end{align}

Fig.~\ref{fig:NRS} shows the model curve using this function along
with the measured data. The parameter $b$ is obtained  by minimizing
the squared error between the modeled rates and measured rates. It
can be seen that the model fits the measured data (e.g., at
$q_{\min}$ and $s_{\max}$) very well.

\subsection{Model for Normalized Rate v.s. Spatial Resolution $\tilde{R}_s(s)$}
$\tilde{R}_s(s)$ is used to describe the reduction
of the normalized bit rate as the frame size reduces. 
As we can see, the desired property for the $\tilde{R}_s(s)$
function is that it should be 1 at $s = s_{\max}$ and monotonically
reduces to 0 at $s = 0$. The data shown in Fig.~\ref{fig:NRS} (first
row) suggest that $\tilde{R}_s(s)$ can be approximated well by a
power function as well, i.e.,
\begin{align}
\tilde{R}_s(s) = \left(\frac{s}{s_{\max}}\right)^c. \label{eq:SCFR}
\end{align}

Fig.~\ref{fig:NRS} shows the model curve using (\ref{eq:SCFR}) along
with the measured data (for all possible $q$ and $t$). The parameter
$c$ is obtained by minimizing the squared error between model predicted rates
and actual measured rates, and characterizes the speed of bit rate
reduction when the frame size decreases.
 It can be seen that the model prediction fits
the actual measurements very well.

\subsection{The Overall Rate Model}
Substituting Eqs. (\ref{eq:SCFR}), (\ref{eq:NRT}) and (\ref{eq:NRQ})
 into (\ref{eq:rstar_abstraction}) leads to the proposed rate model
\begin{align}
  R(q, s, t) = R_{\max}\left(\frac{q}{q_{\min}}\right)^{-a}\left(\frac{t}{t_{\max}}\right)^b\left(\frac{s}{s_{\max}}\right)^c, \label{eq:RSTAR}
\end{align}
where ${q}_{\min}$, $s_{\max}$ and $t_{\max}$ should be chosen based
on the underlying application,  $R_{\max}$ is the rate when coding a
video at $q_{\min}$, $s_{\max}$ and $t_{\max}$, and $a$, $b$ and $c$
are the model parameters, characterizing how fast the rate decreases
when $s$, $t$ reduce and $q$ increases.  Note that $R_{\max}$ is
also a model parameter that is content dependent. Generally a more
complex video (with high motion and complex texture) requires a
higher $R_{\max}$.


\begin{figure*}[htp]
\centering
\includegraphics[scale=0.355]{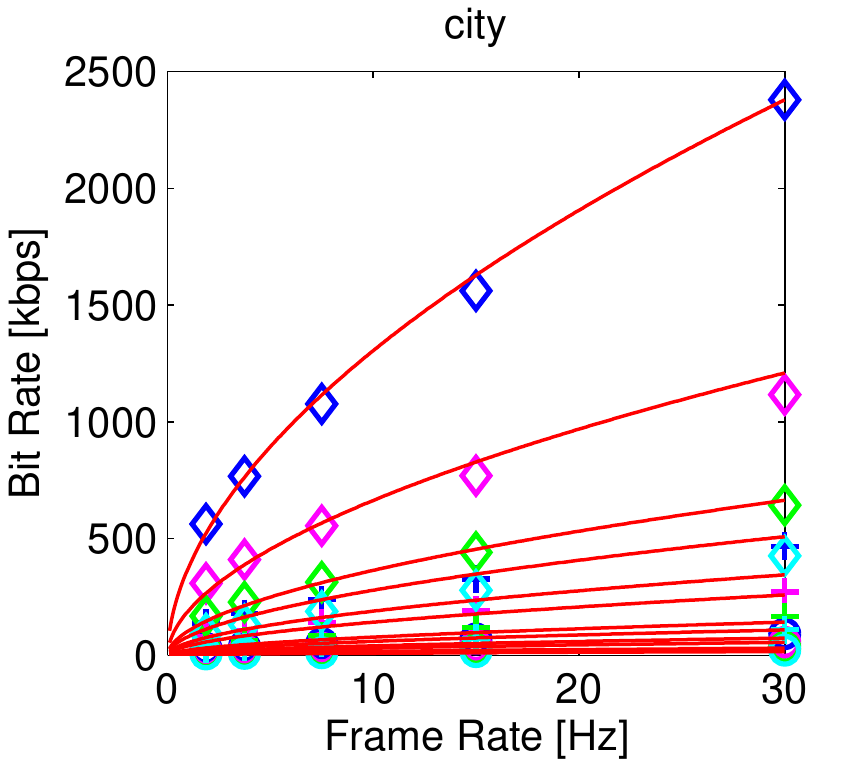}
\includegraphics[scale=0.355]{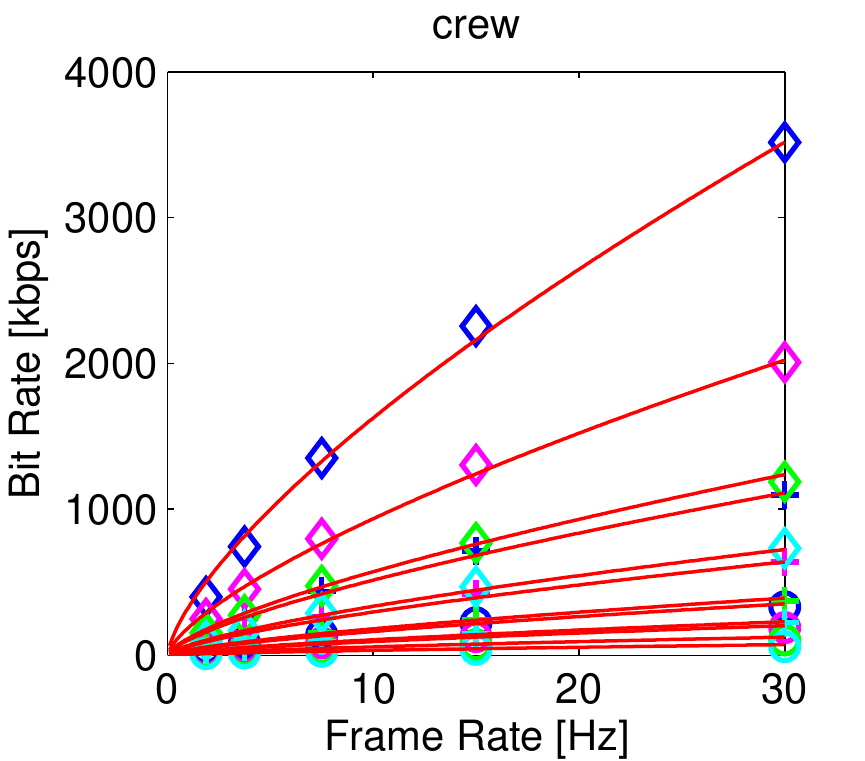}
\includegraphics[scale=0.355]{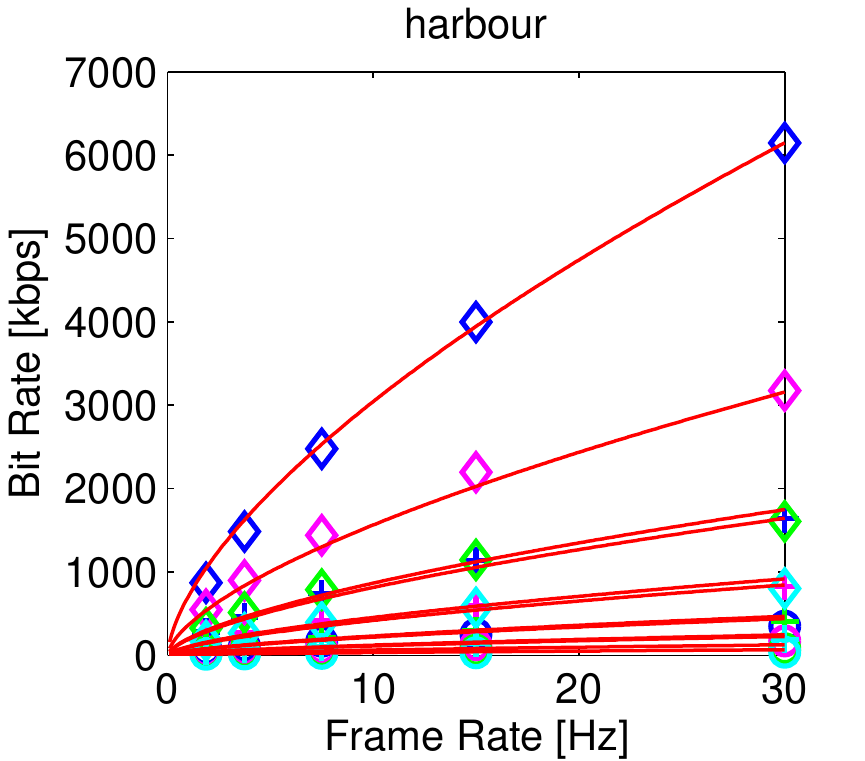}
\includegraphics[scale=0.355]{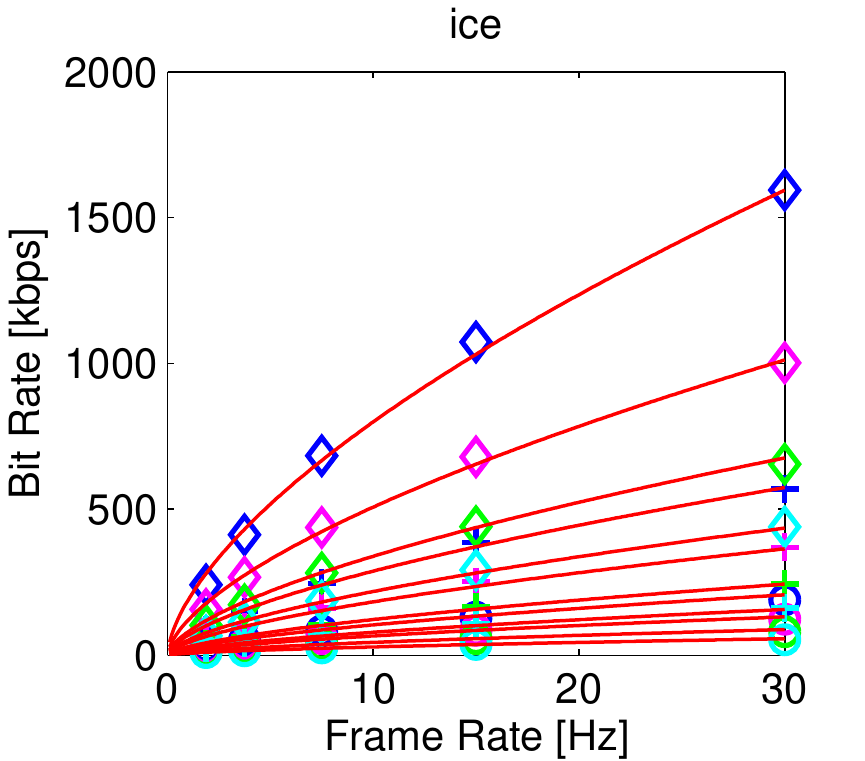}
\includegraphics[scale=0.355]{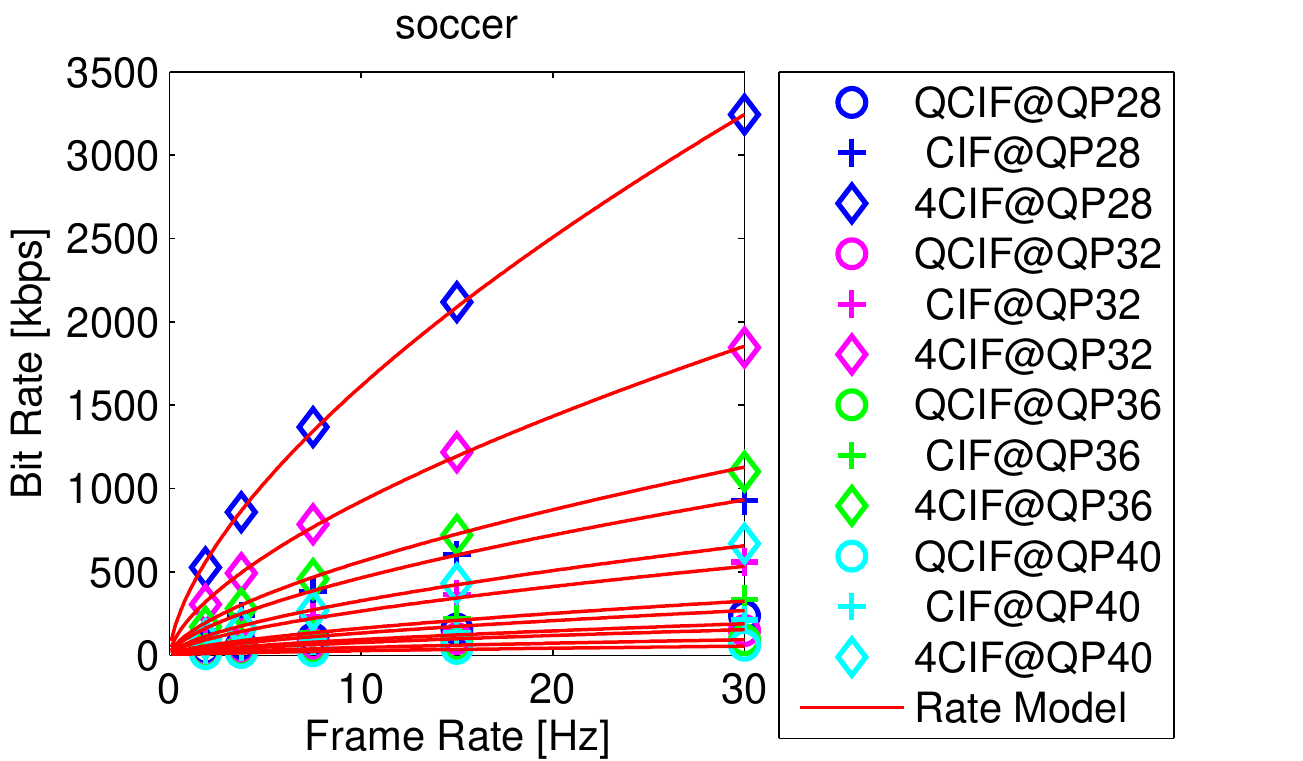}
\caption{Rate prediction using (\ref{eq:RSTAR}) for test sequences at all STAR combinations. }
\label{fig:soccer_RSTAR}
\end{figure*}

Table~\ref{tab:rate_model_accuracy_SVC_eqQP} lists the parameter
values and model accuracy in terms of relative RMSE (i.e., RRMSE =
RMSE/$R_{\max}$), and the Pearson correlation (PC) between measured
and predicted rates, defined as
\begin{align}
r_{xy} = \frac{n\sum{x_{i}y_{i}}-\sum{x_{i}\sum{y_{i}}}}{\sqrt{n\sum{x_{i}^{2}-(\sum{x_{i}})^{2}}}\sqrt{n\sum{y_{i}^{2}-(\sum{y_{i}})^{2}}}},
\end{align} where $x_{i}$ and $y_{i}$ are the measured and predicted rates, and $n$ is the total number of
available samples. We see that the model is very accurate for all
different sequences, with small RRMSE and high PC.
Fig.~\ref{fig:soccer_RSTAR} shows the actual rate data and
corresponding estimated rates for all videos, via the proposed model
(\ref{eq:RSTAR}). Results show that our proposed model can predict
the bit rate very well.



\begin{table}[htp]
\centering
\caption{Rate Model Parameter and Its Accuracy for SVC\#1}
\label{tab:rate_model_accuracy_SVC_eqQP}
\begin{tabular}{|c|c|c|c|c|c|c|}
\hline
&city &crew&harbour&ice&soccer& ave.\\
\hline
$a$ &1.394  &  1.139 &    1.373 &    0.936 &    1.152 & 1.199\\
$b$ &0.547    &0.702   & 0.640   & 0.628 &    0.635 &0.630\\
$c$ &1.114 &    0.830 &    0.952 &    0.736 &    0.899 &0.906\\
$R_{\max}$ & 2379 & 3516 & 6145 & 1594 & 3242 & 3376\\
\hline
RRMSE &1.12\%&    0.75\%&    0.94\%&    0.72\%&    0.41\%&0.80\% \\
PC &0.9985 &  0.9991&    0.9985&    0.9993&    0.9997&0.9990\\
\hline
\end{tabular}
\end{table}

\section{Model Validation}
\label{sec:model_validate}

The model described in the last section was derived based on rate
data obtained for scalable video with joint spatial and temporal
scalability without using QP cascading.  In this section, we verify
that our proposed rate model works for other coding scenarios,
including scalable and single layer video, temporal prediction using
hierarchical B or IPPP, with or without QP cascading, etc. Results
show that our model can accurately predict the bit rates for all
practical coding applications.

\subsection{Model Validation for Video with Joint Spatial and Temporal Scalability}
\label{ssec:model_verify_SVC}
 There are many ways to encode scalable
video in practice. QP cascading is one popular solution. Typically,
we can vary QP in two ways: one is applied on temporal resolution,
where we use smaller QP for pictures at lower temporal layer and
larger QP at higher temporal layer; the other is using relative
smaller QP for lower spatial layer (e.g., base layer) and increasing
the QP along with the spatial resolution increment~\cite{JSVM}.
Currently, SVC reference software -- JSVM uses explicit temporal QP
cascading as its default setting, i.e, $  {\rm QP}_{T} = {\rm
QP}_{0} + 3 + T, T
> 0$,
 where $T$ is the temporal layer identifier. In our simulations, we
apply this default temporal QP cascading without modification. On
the other hand, spatial QP cascading algorithm is not specified
in~\cite{JSVM}. For simplicity, we select two constant delta QPs
between successive spatial layers (noted as dQPs), i.e., dQPs = 4
and dQPs = 6.

 The first three entries in Table~\ref{tab:sim_cases_SVC}\footnote{SVC\#1-3: Joint spatial-temporal scalability with amplitude
 scalability coded using multiple QPs; SVC\#4: combined scalability.} summarizes several cases
 we examined for joint spatial and temporal scalability, where HierB stands for dyadic
hierarchical B prediction structure with the number indicating the
GOP length. \#SR is the number of spatial resolutions (SRs). \#TR is
the number of the temporal layers which can be derived by the GOP
length, i.e., \#TR = $\log_2$GOP + 1. \#AR is the number of
amplitude resolutions (ARs), which is controlled by the QP. The
cited QPs are those used at the base layer, noted as bQP. To provide
multiple amplitude resolutions, SVC codes a video using different
base layer QPs. Please note that Fig.~\ref{fig:soccer_RSTAR} and
Table~\ref{tab:rate_model_accuracy_SVC_eqQP} are the experimental
results for simulation SVC\#1.

\begin{table}[htp]
\scriptsize
  \centering
  \caption{List of Experiments for Scalable Video}
  \label{tab:sim_cases_SVC}
  \begin{tabular}{|c|c|c|c|c|c|}
    \hline
  \multirow{2}{*}{} & \multicolumn{2}{|c|}{QP Cascading} & \multirow{2}{*}{GOP} & \multirow{2}{*}{\#SR} &\multirow{2}{*}{\#AR: bQP}\\   \cline{2-3}
    & temporal & spatial &  &&\\
    \hline
    SVC\#1 & NO   & NO  & HierB: 16 &                           3 &4: 28, 32, 36, 40\\
    SVC\#2 & Yes & Yes: dQPs = 4 & HierB: 16 & 3 &3: 16, 20, 24\\
    SVC\#3 & Yes & Yes: dQPs = 6 & HierB: 8  & 3 &4:  24, 28, 32, 36\\
    SVC\#4 & Yes & Yes: dQPs = 6 & HierB: 8  & 2 &3: 16, 20, 24 \\
    \hline
   \end{tabular}
\end{table}

\begin{table}[htp]
\centering
\caption{Rate Model Parameter and Its Accuracy For SVC\#2}
\label{tab:rate_model_accuracy_SVC_dQP4}
\begin{tabular}{|c|c|c|c|c|c|c|}
\hline
&city &crew&harbour&ice&soccer& ave.\\
\hline
$a$ &1.342  &  1.20 &    1.171 &    0.952 &    1.092 &1.151\\
$b$ &0.329    &0.538   & 0.508   & 0.496 &    0.454 &0.465\\
$c$ &0.806 &    0.533 &    0.646 &    0.537 &    0.642 &0.633\\
$R_{\max}$ & 3625 &  4960&   8675 &  2334 &  4554& 6037\\
\hline
RRMSE &1.03\%&    1.26\%&    1.60\%&    1.19\%&    1.14\%& 1.24\% \\
PC &0.9985 &  0.9974&    0.9956&    0.9979&    0.9980& 0.9975\\
\hline
\end{tabular}
\end{table}

Table~\ref{tab:rate_model_accuracy_SVC_dQP4} and
~\ref{tab:rate_model_accuracy_SVC_dQP6} present the prediction
accuracy and model parameters for  SVC\#2 and SVC\#3 respectively.
According to the simulations, we can see that our proposed rate
model is generally applicable regardless the coding structures. We
can also notice that model parameters are highly content dependent,
and their values depend on coding scenarios as well.

\begin{table}[htp]
\centering
\caption{Rate Model Parameter and Its Accuracy For SVC\#3}
\label{tab:rate_model_accuracy_SVC_dQP6}
\begin{tabular}{|c|c|c|c|c|c|c|}
\hline
&city &crew&harbour&ice&soccer& ave.\\
\hline
$a$ &1.239  &  1.092 &    1.363 &    0.953 &    1.15 &1.159\\
$b$ &0.268    &0.459   & 0.288   & 0.447 &    0.425 &0.377\\
$c$ &0.512 &    0.319 &    0.427 &    0.371 &    0.411 &0.408\\
$R_{\max}$ & 761 & 1169 & 1953& 761&1200&1169\\
\hline
RRMSE &1.68\%&   0.96\%&    2.09\%&    1.14\%&    1.59\%&   1.49\%\\
PC &0.9963 &  0.9986&    0.9942&    0.9980&    0.9962 & 0.9967\\
\hline
\end{tabular}
\end{table}

\subsection{Model Validation for Combined Spatial, Temporal and Amplitude Scalability} \label{sec:model_verify_combined}
\begin{figure}[htp]
\centering
\includegraphics[scale=0.45]{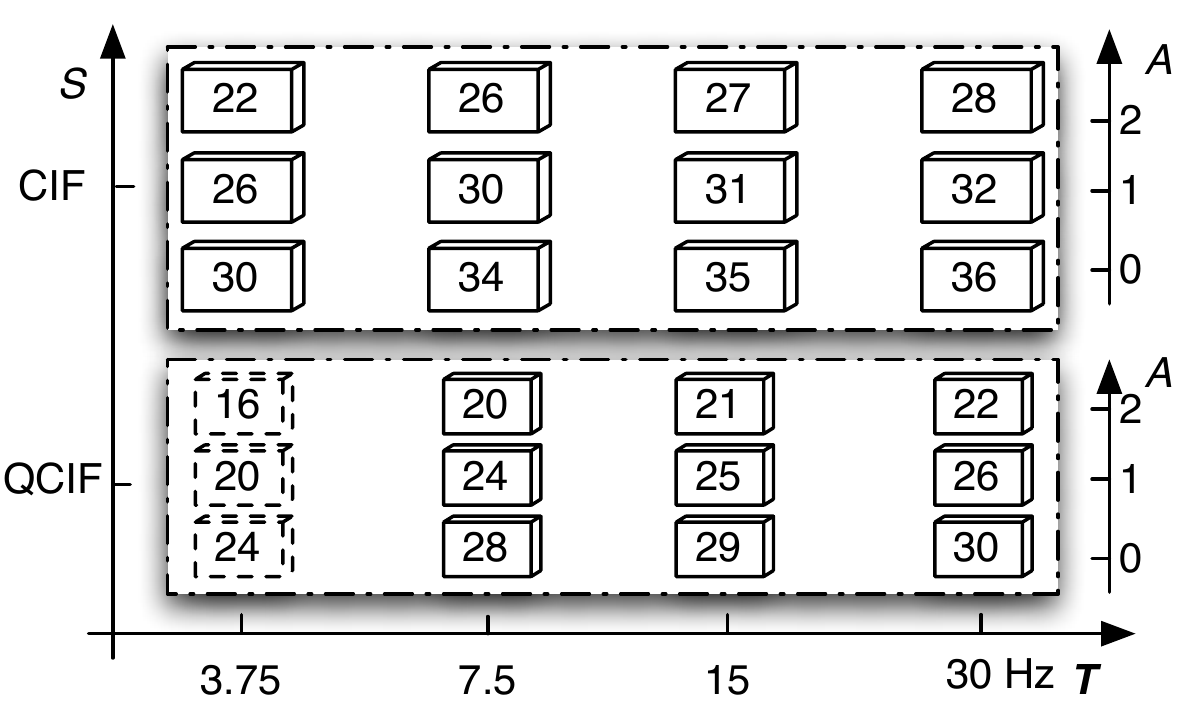}
\caption{Illustrative layered structure for SVC\#4:  $A=0$ is the amplitude base layer.
Different QPs are applied to temporal/spatial enhancement layers by enabling QP cascading.
Delta QP is 4 and 6 for successive amplitude and spatial layers, respectively.}
\label{fig:combS}
\end{figure}

SVC\#4 in Table~\ref{tab:sim_cases_SVC} refers to combined
scalability of SVC, which can provide different STAR combinations
within a single scalable stream. SVC\#4 uses two spatial layers,
three amplitude layers and four temporal layers as shown in
Fig.~\ref{fig:combS}. QP cascading is used and actual QP used in
different layers are illustrated in Fig.~\ref{fig:combS}.

Table~\ref{tab:rate_model_accuracy_SVC_4}
present the results for model
parameters and prediction accuracy. It shows the same model also works
well for the combined scalability.

\begin{table}[htp]
\centering
\caption{Rate Model Parameter and Its Accuracy For SVC\#4}
\label{tab:rate_model_accuracy_SVC_4}
\begin{tabular}{|c|c|c|c|c|c|c|}
\hline
&city &crew&harbour&ice&soccer& ave.\\
\hline
$a$ &0.881  &  0.69 &    0.768 &    0.647 &    0.771 &0.751\\
$b$ &0.254  & 0.536   & 0.471   & 0.486 &    0.441 &0.438\\
$c$ &0.902 &    0.605 &    0.808 &    0.669 &    0.799 &0.757\\
$R_{\max}$ &1816&2909&4556&1518&2588&2678\\
\hline
RRMSE &2.19\%&   2.67\%&    1.89\%&    2.24\%&    1.52\%&   2.10\%\\
PC &0.9968 &  0.9942&    0.9971&    0.9962&    0.9983 & 0.9965\\
\hline
\end{tabular}
\end{table}

\subsection{Model Validation for Single Layer Video}
\label{ssec:model_verify_SL} In this section, we validate our
proposed rate model for singe layer video encoding.  Video sequences
are encoded with different combinations of FS, FR and QS, using
JSVM~\cite{JSVM} single layer mode. To code a video at different
SRs, we first down-sample the original video to the desired SR using
the filter suggested by~\cite{JSVMEncoderDescp}, and then code the
video at that SR.  Table~\ref{tab:sim_cases_SL} summarizes the three
settings we examined. For SL\#2 and SL\#3, multiple temporal
resolutions (TRs) are obtained using the HierB structure; whereas
for SL\#1, each TR is obtained by temporally down-sampling the
original video to the desired TR. Different QPs are used to provide
multiple amplitude resolutions.
Table~\ref{tab:rate_model_accuracy_SL_1}-~\ref{tab:rate_model_accuracy_SL_3}
present the prediction accuracy and model parameters for
SL\#1-SL\#3. In summary, our rate model can predict the bit rates
accurately for different single layer coding scenarios.

\begin{table}[htp]
  \centering
  \caption{Experiments for Singe Layer Video}
  \label{tab:sim_cases_SL}
  \begin{tabular}{|c|c|c|c|c|c|}
    \hline
  \multirow{2}{*}{} & \multicolumn{2}{|c|}{QP Cascading} & \multirow{2}{*}{GOP}& \multirow{2}{*}{\#AR: bQP} \\  \cline{2-3}
    & temporal & spatial &  &\\
    \hline
    SL\#1 & NO  & NO & IPPP:8  &  4: 24, 28, 32, 36 \\
    SL\#2 & Yes & NO & HierB:8  & 4: 28, 32, 36, 40 \\
    SL\#3 & Yes & Yes: dQPs = 6    &HierB:16 &  3: 16, 20, 24\\
    \hline
   \end{tabular}
\end{table}

\begin{table}[htp]
\centering
\caption{Rate Model Parameter and Its Accuracy for SL\#1}
\label{tab:rate_model_accuracy_SL_1}
\begin{tabular}{|c|c|c|c|c|c|c|}
\hline
&city &crew&harbour&ice&soccer& ave.\\
\hline
$a$ &1.935  &  1.362 &    1.23 &    1.12 &    1.38 &1.405\\
$b$ &0.836    &0.828   & 0.795   & 0.679 &    0.711 &0.770\\
$c$ &1.301 &    0.881 &    0.895 &    0.729 &    0.992 &0.960\\
$R_{\max}$ & 7561 &  6962 & 10884 & 2140 & 6084&6727\\
\hline
RRMSE &0.76\%&   0.93\%&    0.95\%&    1.15\%&    0.81\%&  0.92\%  \\
PC &0.9992 &  0.9988&    0.9983&    0.9983&    0.9991 & 0.9987\\
\hline
\end{tabular}
\end{table}

\begin{table}[htp]
\centering
\caption{Rate Model Parameter and Its Accuracy for SL\#2}
\label{tab:rate_model_accuracy_SL_2}
\begin{tabular}{|c|c|c|c|c|c|c|}
\hline
&city &crew&harbour&ice&soccer& ave.\\
\hline
$a$ & 1.371 &  1.095 &    1.248 &   0.86  &    1.086 &1.132\\
$b$ &  0.233 &0.471   & 0.397   & 0.438 &    0.39 &0.386\\
$c$ & 1.047 &    0.785 &    0.894 & 0.667   &    0.88 &0.855\\
$R_{\max}$ &1512 &   2429 &  3818 &  975 &  2268 &2201\\
\hline
RRMSE & 0.70\%&   0.66\%&    1.23\%& 0.72\% &    0.65\%& 0.79\%  \\
PC & 0.9995&  0.9993&    0.9977&   0.9992 &    0.9995 & 0.9990\\
\hline
\end{tabular}
\end{table}

\begin{table}[htp]
\centering
\caption{Rate Model Parameter and Its Accuracy for SL\#3}
\label{tab:rate_model_accuracy_SL_3}
\begin{tabular}{|c|c|c|c|c|c|c|}
\hline
&city &crew&harbour&ice&soccer& ave.\\
\hline
$a$ & 1.333 &  1.054 &    1.149 &   0.851  &    1.037 &1.085\\
$b$ &  0.242 &0.491   & 0.422   & 0.454 &    0.403 &0.402\\
$c$ & 0.479 &    0.266 &    0.361 & 0.239   &    0.40 &0.349\\
$R_{\max}$ &1965 &2969&4909&1125&2736&2741\\
\hline
RRMSE & 1.26\%&   1.24\%&    1.98\%& 1.19\% &    1.19\%&  1.37\% \\
PC & 0.9974&  0.9970&    0.9924 &   0.9971 &    0.9975 & 0.9963\\
\hline
\end{tabular}
\end{table}


\section{Model Parameter Prediction Using Content Features} \label{sec:para_pred}
As shown in previous sections,  model parameters are highly content dependent.
In this section, we investigate
how to predict the parameters accurately using content features which can be easily obtained
from original video signals. We have four parameters in total for our rate model,
i.e., $a$, $b$, $c$ and $R_{\max}$.

\begin{figure}[htp]
\centering
\subfigure[SVC\#1]{\includegraphics[scale=0.4]{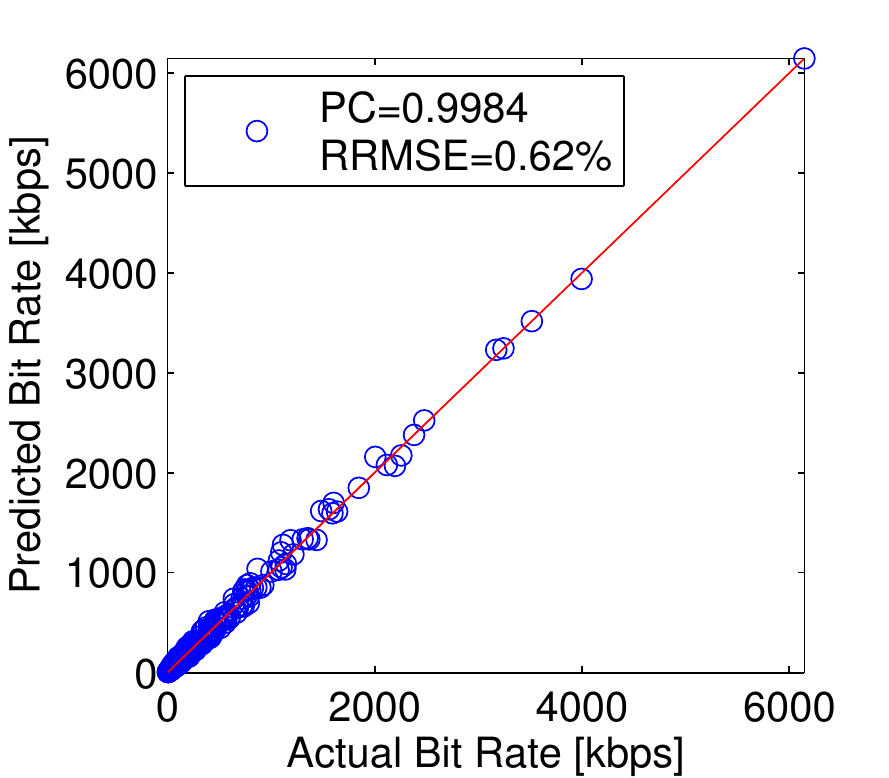}}
\subfigure[SL\#2]{\includegraphics[scale=0.4]{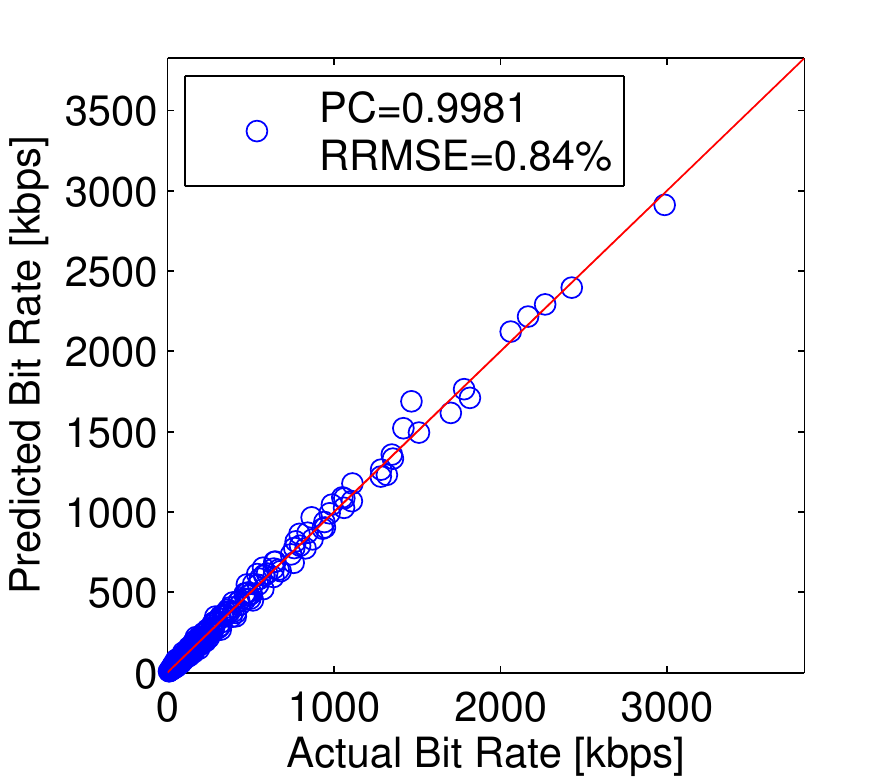}}
\caption{Actual measured versus model predicted bit rates for all test sequences under SVC\#1 and SL\#2,
where model parameters are predicted using content features. Other coding scenarios have the similar
high performance for bit rate estimation using (\ref{eq:RSTAR}) with content predicted parameters. }
\label{fig:model_para_pred}
\end{figure}

To predict these parameters, we adopt the same approach presented
in~\cite{Ma_RateQualityModel}, which predicts five parameters for
both rate and quality models (three for the rate model, and two for
the quality model) considering only the impact of FR and QS. In a
nut shell, we predict each parameter as a weighted combination of
some chosen features plus a constant, and determine the best set of
features and corresponding weights using a cross-validation
criterion. We have found that the same set of three features,
$\mu_{\tt DFD}$, $\sigma_{\tt MVM}$ and $\sigma_{\tt MDA}$,  can
predict the model parameters under different coding scenarios. These
features stand for the mean of the displaced frame difference,
standard deviation of the motion vector magnitude and standard
deviation of the motion direction activity, respectively. The
predictor can be written as
\begin{align}
\bf P = H F, \label{eq:para_pred}
\end{align} where  ${\bf F} = [1, \mu_{\tt DFD}, \sigma_{\tt MVM}, \sigma_{\tt MDA}]^T$, ${\bf P}= \left[a, b, c,
R_{\max}\right]^T$. The predictor matrix $\bf H$ depends on the
coding structure. For example, for SVC\#1 and SL\#2, they are
\begin{equation}
{\bf H}_{\tt SVC\#1} = \left[ \begin{array}{cccc}

1.374   &0.059  &-0.049 &-0.253 \\
0.226   &0.022  &-0.007 &0.305  \\
1.507   &0.005  & 0.0013    &-0.594 \\
-7262   &1240   &-995.0 &8033  \\

\end{array} \right],\label{eq:H}
\end{equation}
and
\begin{equation}
{\bf H}_{\tt SL\#2} = \left[ \begin{array}{cccc}

1.538   &0.040  &-0.025 &-0.474\\
-0.241  &0.025  &-0.014  &0.530\\
1.420   &0.011  &0.0099 &-0.619\\
-4598   &795.9  &-549.2 &4810\\

\end{array} \right]. \label{eq:H}
\end{equation}   Please
refer to~\cite{Ma_RateQualityModel, zhan_thesis} for more details.
Fig.~\ref{fig:model_para_pred} shows the model accuracy using
(\ref{eq:RSTAR}) for SVC\#1 and SL\#2 with parameters predicted
using content features.

\section{Application} \label{sec:svc_adapt}

In traditional encoder rate-control algorithms, the spatial and
temporal resolutions are pre-fixed based on some empirical rules,
and the encoder varies the QS, to reach a target bit rate. Selection
of QS is typically based on models of rate versus QS. When varying
the QS alone cannot meet the target bit rate, frames are skipped as
necessary. Joint decision of QS and frame skip has also been
considered, but often governed by heuristic rules, or using the mean
square error (MSE)~\cite{Kuo_JointSpatialTemporalRC} as a quality
measure. Ideally, the encoder should choose STAR that leads to the
best perceptual quality, while meeting the target bit rate.

In video streaming, the same video is often requested by receivers
with diverse sustainable receiving rates. To address this diversity,
a video may be coded into a scalable stream that can be decoded at
different STARs. Given a particular user's sustainable rate, either
the sender or a transcoder at a proxy needs to extract from the
original bitstream at certain layers   that together corresponds to
a certain STAR to meet the rate constraint. This problem is
generally known as scalable video adaptation. Here again the
challenging problem is to determine which layers (which corresponds
to a particular STAR) to extract, to maximize the perceptual
quality.

Another related problem is how to order the temporal, spatial and
amplitude layers of a SVC bitstream into a rate-quality optimized
layered stream, so that each additional layer leads  to  the maximum
ratio of the quality gain over the rate increment,  and yet the
decoding of each new layer only depends on this layer and previous
layers.

Obviously, the solution of the above problems requires accurate rate
and quality models that can predict the rate and quality associated
with a given STAR. In this section, we discuss solutions to the
above problems using both the rate model presented here and the
QSTAR model described in~\cite{QSTAR}, which relates the perceived
quality with the STAR.

Our prior work in \cite{Ma_RateQualityModel, Hao_MM} have studied
similar problems but considering only the optimization of FR and QS,
assuming the FS is fixed. Here, we extend these studies to consider
the adaptation of FS, FR and QS jointly.  We first review the QSTAR
quality model. We then describe how to optimize the STAR in
rate-constrained video coding and adaptation.  Finally we  present
two algorithms for ordering the SVC layers.

\subsection{Analytical Quality Model}
The QSTAR model proposed in~\cite{QSTAR} relates the quality with
$(q, s, t)$ by
\begin{align}
  &Q(q, s, t) = \nonumber\\ &\mbox{}\frac{1-e^{-\alpha_q\left(\frac{q_{\min}}{q}\right)^{\beta_q}}}{1-e^{-\alpha_q}}\frac{1-e^{-\alpha_s(q)(\frac{s}{s_{\max}})^{\beta_s}}}{1-e^{-\alpha_s}}\frac{1-e^{-\alpha_t(\frac{t}{t_{\max}})^{\beta_t}}}{1-e^{-\alpha_t}},\label{eq:QSTAR}
\end{align} where $\beta_q = 1$, $\beta_s$ = 0.74,
$\beta_t$ = 0.63, $\alpha_s(q) = \tilde{\alpha}_s(\nu_1 {\rm QP}(q)
+ \nu_2) $ when QP $\geq 28$ and $\alpha_s = \tilde{\alpha}_s(28
\nu_1 + \nu_2) $ when QP $<$ 28, with $\nu_1 = -0.037, \nu_2 =
2.25$. The rate-model parameters, $a$, $b$, $c$, $R_{\max}$ and
quality model parameters $\alpha_q$, $\hat{\alpha}_s$ and $\alpha_t$
control how fast the rate and quality, respectively, drop as the
spatial, temporal, or amplitude resolution decreases. These
parameters depend on the the motion and texture characteristics of
the underlying video,  and can be estimated from certain features
computable from original video (as described in
Section~\ref{sec:para_pred}).
Table~\ref{tab:quality_model_accuracy_SVC_adapt} shows the quality
model parameters and its accuracy,  for the same set of test
sequences used for the rate model development. Although the study
in~\cite{QSTAR} used scalable video coded with H.264/SVC
codec~\cite{JSVM}, a separate study~\cite{ICIP_SVC_AVC_COMP} has
confirmed that there is no statistically significant quality
difference between scalable (coded using H.264/SVC) and non-scalable
video (coded using H.264/AVC) when coded at the same STAR. This
means that the QSTAR is applicable to both scalable and non-scalable
video, and the model parameter is generally independent of the
encoder setting.

\begin{table}[htp]
\footnotesize \centering \caption{Quality Model Parameter and Its
Accuracy} \label{tab:quality_model_accuracy_SVC_adapt}
\begin{tabular}{|c|c|c|c|c|c|c|}
\hline
&city &crew&harbour&ice&soccer& ave.\\
\hline
$\alpha_q$ &7.25  &  4.51 &    9.65 &    5.61 &    6.31 &6.67\\
$\hat{\alpha}_s$ &3.52    &4.07   & 4.58   & 3.68 &    4.55 &4.08\\
$\alpha_t$ &4.10 &    3.09 &    2.83 &    3.00 &    2.23 &3.05\\
\hline
RRMSE &1.80\%&    2.50\%&    3.80\%&    3.30\%&    3.20\%& 2.92\% \\
PC &0.998 &  0.996&    0.992&    0.993&    0.992&0.995\\
\hline
\end{tabular}
\end{table}

\subsection{Rate Constrained Quality and STAR Optimization}
Although video encoding and adaptation are quite different
applications, the essence of these problem is to maximize the video
quality under the bit rate constraint, i.e.,
\begin{align}
&{\mbox{Determine $q, s, t$ to~}}{\mbox{maximize~~}}  Q(q, s, t) \nonumber \\
          &{\mbox{{~~~~~~~~~~~~~~~~~~~~~~~~subject to~~}}} R(q, s, t) \leq R_0, \label{eq:adapt_prob}
\end{align} where $R_0$ is the bit rate constraint.
We  employ the rate and quality models in \eqref{eq:RSTAR} and
\eqref{eq:QSTAR} to solve this optimization problem, first assuming
$s$, $t$, and $q$ can take on any value in a continuous range, $s\in
(0, s_{\max}]$, $t\in(0, t_{\max}]$, and $q\in[q_{\min}, \infty)$.
We then describe the solution when $s$ and $t$ are chosen from
discrete sets feasible with dyadic temporal and spatial prediction
structures typically adopted by  practical encoders.

\subsubsection{Optimal solution assuming continuous $s$, $t$, $q$  and Quality-Rate Model}

\begin{figure}
\centering
  \includegraphics[scale=0.34]{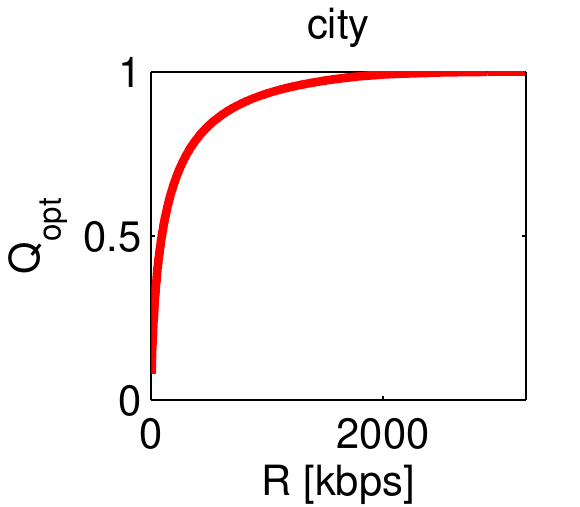}
  \includegraphics[scale=0.34]{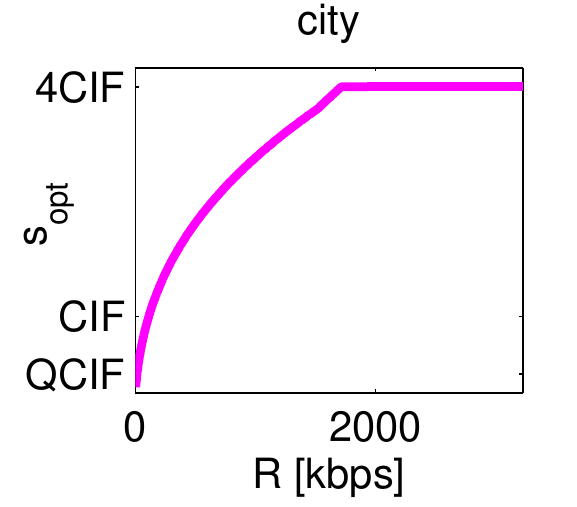}
  \includegraphics[scale=0.34]{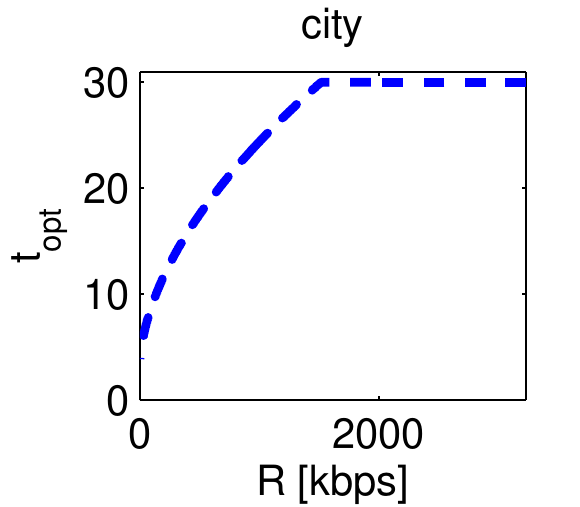}
  \includegraphics[scale=0.34]{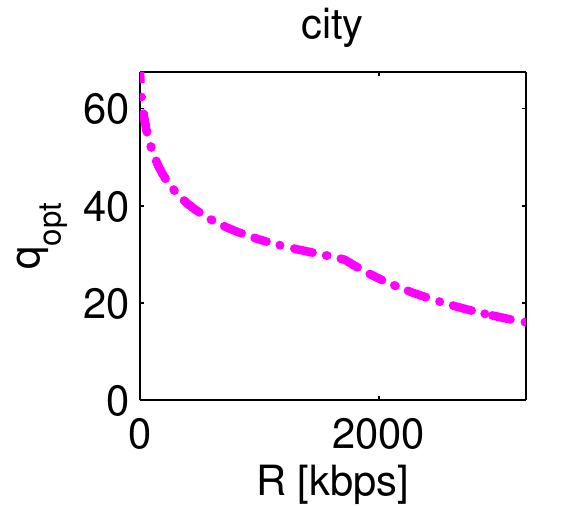}

  \includegraphics[scale=0.34]{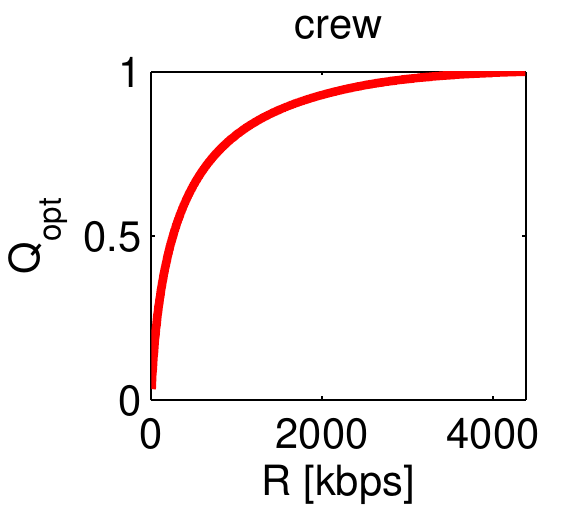}
  \includegraphics[scale=0.34]{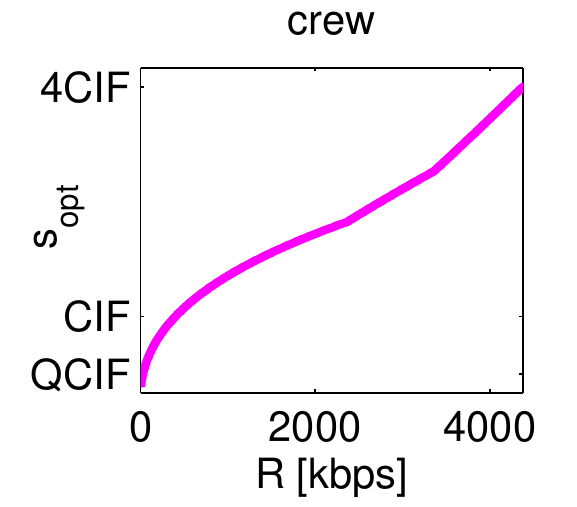}
  \includegraphics[scale=0.34]{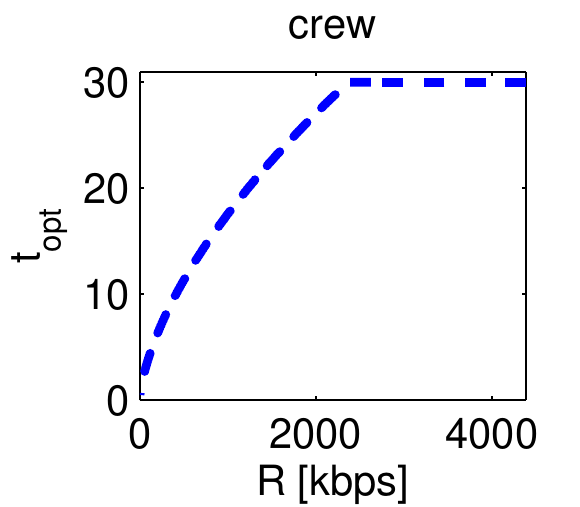}
  \includegraphics[scale=0.34]{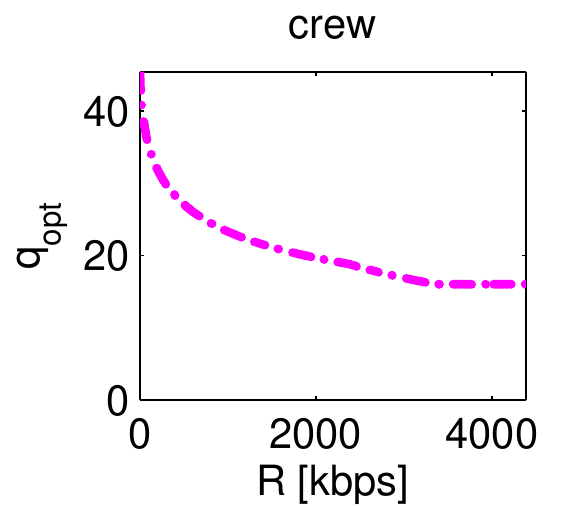}

  \includegraphics[scale=0.34]{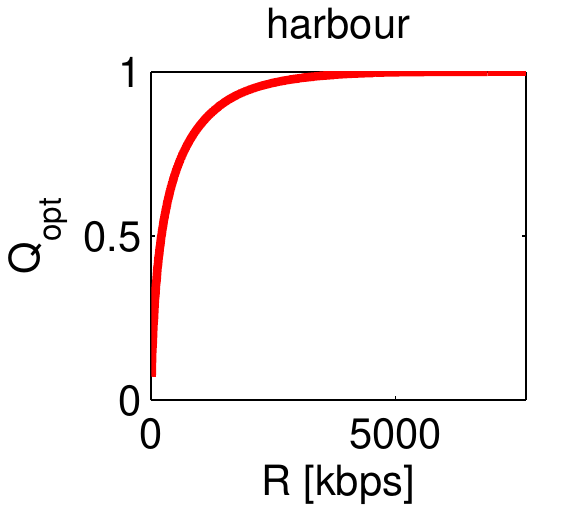}
  \includegraphics[scale=0.34]{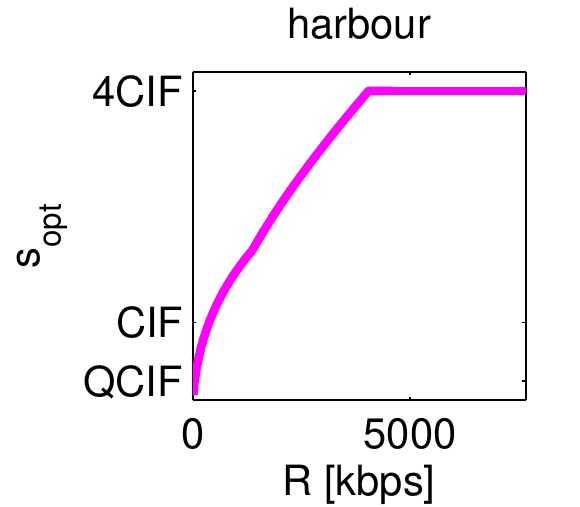}
  \includegraphics[scale=0.34]{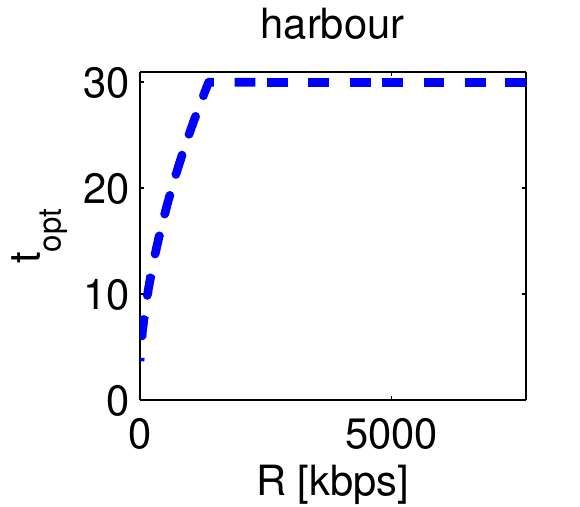}
  \includegraphics[scale=0.34]{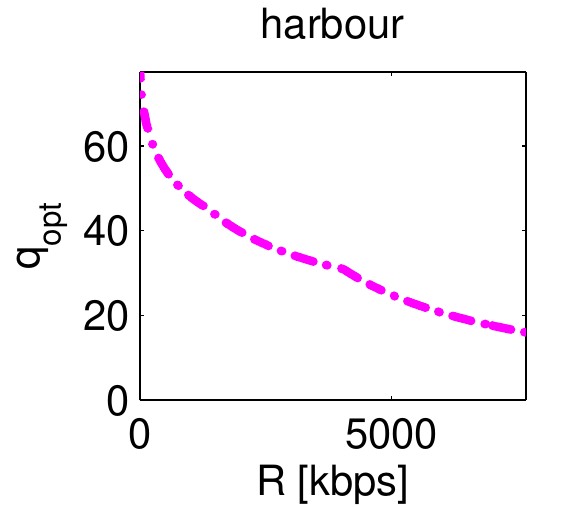}

  \includegraphics[scale=0.34]{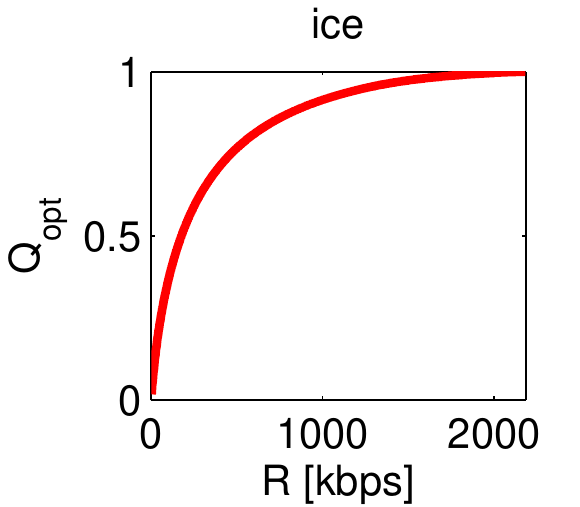}
  \includegraphics[scale=0.34]{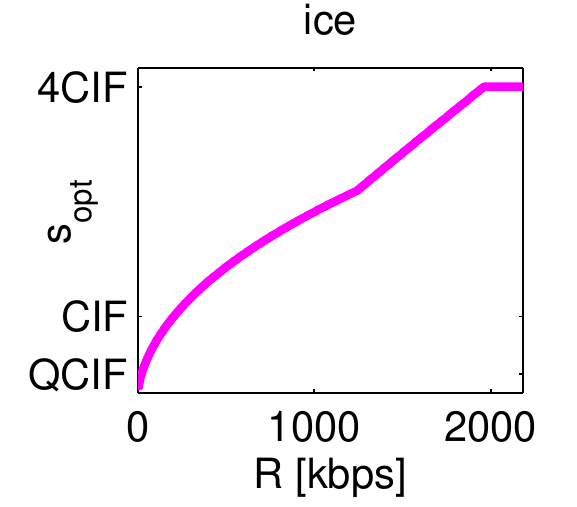}
  \includegraphics[scale=0.34]{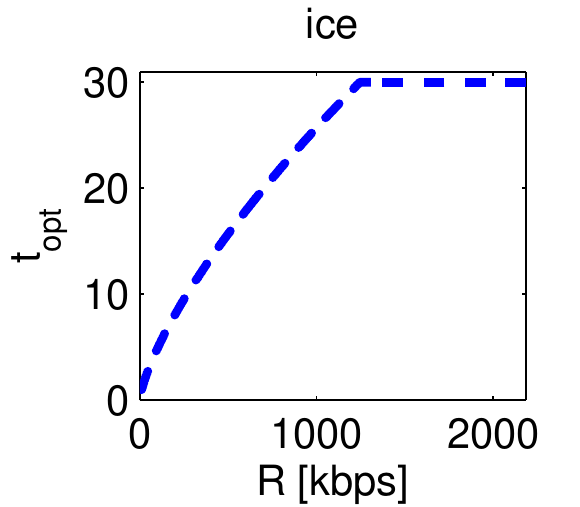}
  \includegraphics[scale=0.34]{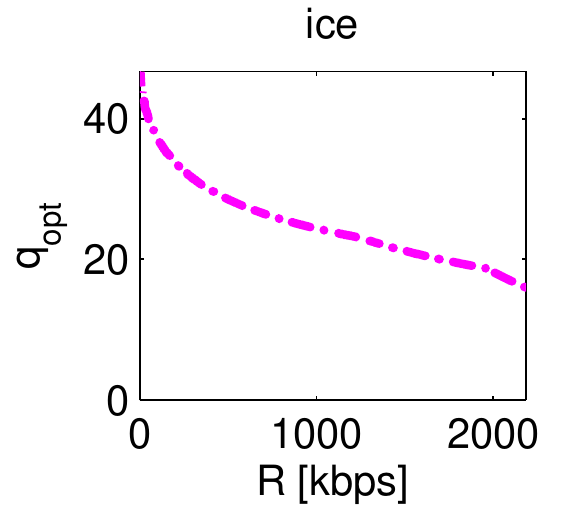}

  \includegraphics[scale=0.34]{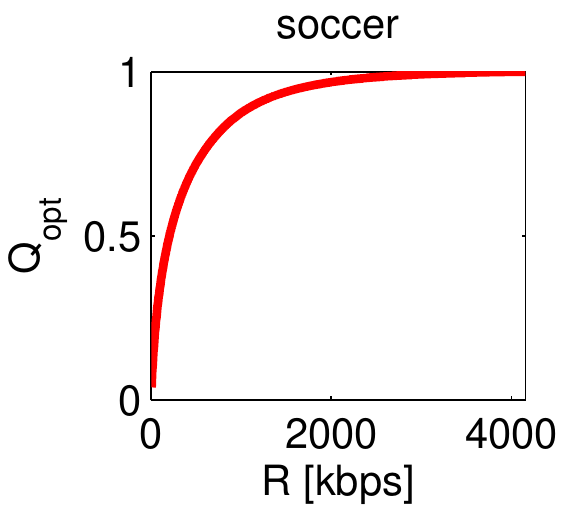}
  \includegraphics[scale=0.34]{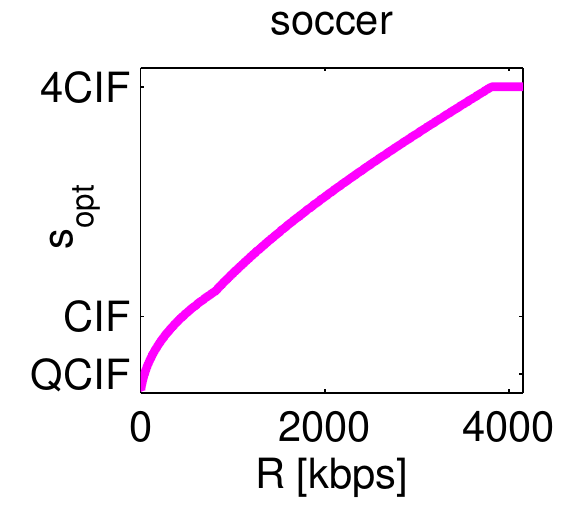}
  \includegraphics[scale=0.34]{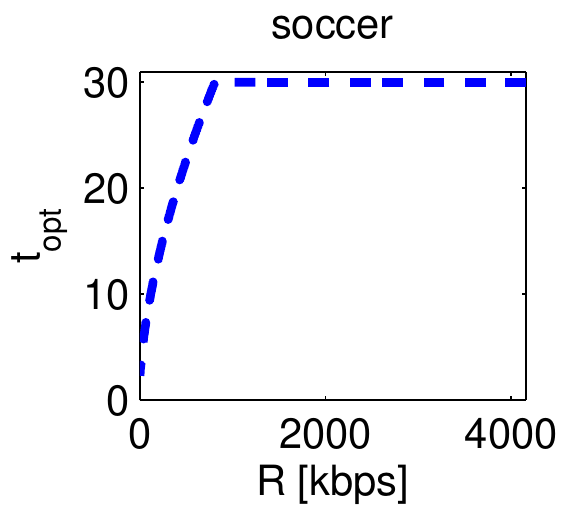}
  \includegraphics[scale=0.34]{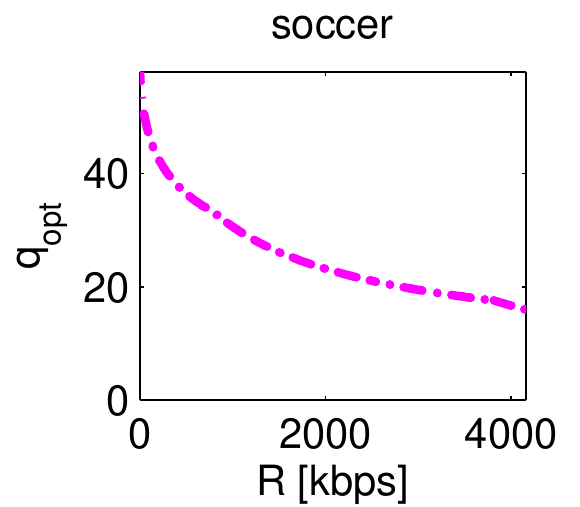}

  \caption{Optimal $Q$, $s$, $t$ and $q$ versus $R$
  by assuming continuous $q$, $s$ and $t$.}
  \label{fig:RQSTAR_continuous}
  \vspace{-0.1in}
\end{figure}

Letting $R(q,s,t) = R_0$ in (\ref{eq:RSTAR}), we obtain
\begin{align}
  q = q_{\min}\sqrt[a]{\left(\frac{R_{\max}}{R_0}\right)\cdot{\left(\frac{s}{s_{\max}}\right)}^{c}\cdot{\left(\frac{t}{t_{\max}}\right)}^{b}}, \label{eq:q_func}
\end{align} which describes the feasible $q$ for a given pair of $s$ and $t$, to satisfy the rate
constraint $R_0$. Substituting \eqref{eq:q_func} into
\eqref{eq:QSTAR} yields the quality model with respect to FS and FR,
i.e., $Q(s, t)$. By solving $\partial{Q({s}, {t})}/\partial{{s}} =
0$, and $\partial{Q({s}, {t})}/\partial{{t}} = 0$, we can have the
optimal pair of ${s}_{\tt opt}$ and ${t}_{\tt opt}$ , and
correspondingly $q_{\tt opt}$ via \eqref{eq:q_func}, to produce the
best quality $Q_{\tt opt}$. However, it is difficult to derive the
close form of $Q_{\tt opt}(R)$, $s_{\tt opt}(R)$, $t_{\tt opt}(R)$
and $q_{\tt opt}(R)$. Thus, we solve the optimization problem
(\ref{eq:adapt_prob}) numerically, by searching over a discrete
space of $(s,t)$, i.e., for any given rate, we search through
feasible $s$ and $t$, derive the $q$ via \eqref{eq:q_func} and
select the $(s,t)$ and consequently $q(s,t)$ that yields the best
quality. Fig.~\ref{fig:RQSTAR_continuous} shows $Q_{\tt opt}$,
$q_{\tt opt}$, $s_{\tt opt}$ and $t_{\tt opt}$ as functions of the
rate constraint $R_0$. As expected, as the rate increases, $s_{\tt
opt}$ and $t_{\tt opt}$ increase while $q_{\tt opt}$ reduces, and
the achievable best quality continuously improve. In this example,
we use the rate model parameters derived for scalable video coded
using SVC\#1 setting, with model  parameters given in
Table~\ref{tab:quality_model_accuracy_SVC_adapt}. The same
methodology can be used both for choosing optimal STAR in a single
layer encoder, or choosing the STAR to extract a scalable
 stream  using different encoder settings, for given target rate, but we would need to use corresponding
 rate model parameters.

\begin{table}[htp]
\vspace{-0.1in}
\centering \caption{$Q(R)$ model parameter and accuracy }
\label{tab:QR_approx}
\begin{tabular}{|c|c|c|c|c|c|}
  \hline
  & city & crew & harbour & ice & soccer \\
  \hline
  $\kappa$ & 5.058&  3.121 & 5.882 & 2.769 & 4.103 \\
  RMSE & 0.49\% &    0.13\%& 0.43\% & 1.3\% & 0.79\%\\
  \hline
\end{tabular}
\end{table}

Although it is difficult to derive the closed form for the $Q_{\tt
opt}(R)$ function, we found that the Q-R relation in
Fig.~\ref{fig:RQSTAR_continuous} can be well approximated by an
inverted exponential function of the form
\begin{align}
Q(R) = \frac{1- e^{-\kappa\left(\frac{R}{R_{\max}}\right)^{0.55}}}{1-e^{-\kappa}}, \label{eq:QR}
\end{align} with $\kappa$ as the model parameter. Table~\ref{tab:QR_approx}
shows the parameter $\kappa$ and approximation accuracy in terms of the root mean squared error (RMSE).

\subsubsection{Optimal solution assuming dyadic $s$ and $t$}
\label{ssec:discrete_s_t} In practical video encoder,  $t$ and $s$
only take on limited discrete values. Here, we consider the popular
dyadic temporal and spatial prediction structure in which  $t$
doubles in increasing temporal  layers, whereas $s$ quadruples  in
increasing spatial layers. We further assume $t_{\tt max} = 30$ and
$s_{\tt max} =$ 4CIF,  so that  $t\in\{3.75, 7.5, 15, 30\}$, and
$s\in\{{\rm QCIF, CIF, 4CIF}\}$. We further assume $q\in[16, 104]$.
These are the ranges in which the original rate and quality models
are derived.

To obtain the optimal solution under this scenario, for each given
rate, we search through all possible $(s_k,t_k)$ pairs from the
feasible sets and and their corresponding $q_k=q(s_k,t_k)$ values
using (\ref{eq:q_func}), and select the one that
 leads to the highest quality. The results are
shown in Fig. \ref{fig:RQSTAR_discrete}. Because the frame rate and
frame size can only increase in discrete steps, the optimal $q$ does
not decrease monotonically with the rate. Rather, whenever either
$s_{\tt opt}$ or $t_{\tt opt}$ jumps to the next higher value,
$q_{\tt opt}$ first increases to meet the rate constraint, and then
decreases while $s$ and/or $t$ is held constant, as the rate
increases.  Also,  when $s$ jumps to the next high level, $t$ may
first decrease  to a  lower level, for example, ICE around 750 kbps.
 Note that the rates at which $s$ jumps
or $t$ jumps are sequence dependent. For sequences with high texture
details (e.g., city), we see that $s$ jumps to the highest level
earlier. For sequences with high motion (e.g., soccer), $t$ jumps to
the highest level earlier and stay at that level even after $s$
jumps.

\subsection{Quality Optimized Layer Ordering for SVC Bit streams}
\label{sec:SVC_order}

\begin{figure}
\centering
  \includegraphics[scale=0.34]{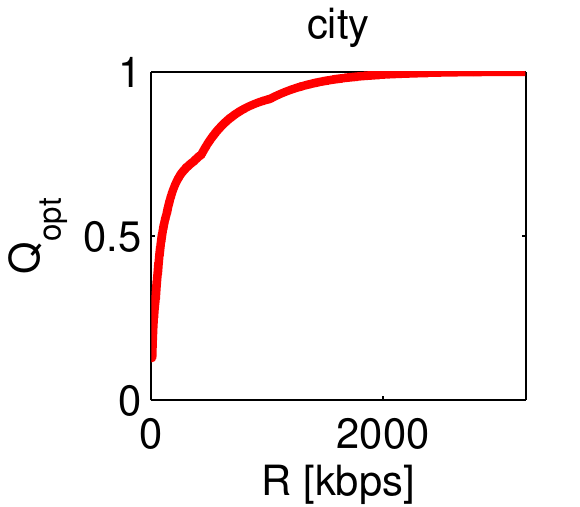}
  \includegraphics[scale=0.34]{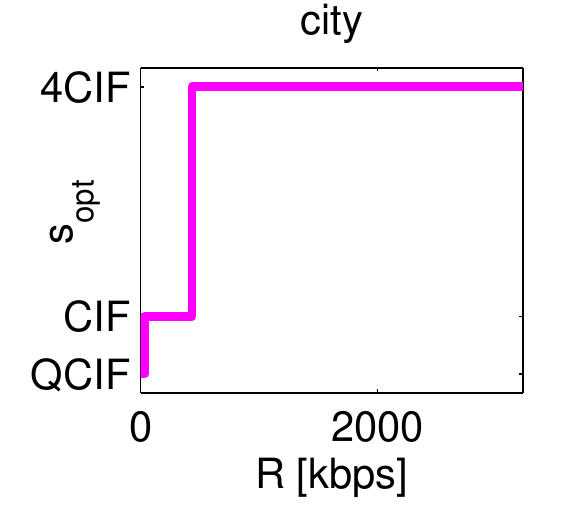}
  \includegraphics[scale=0.34]{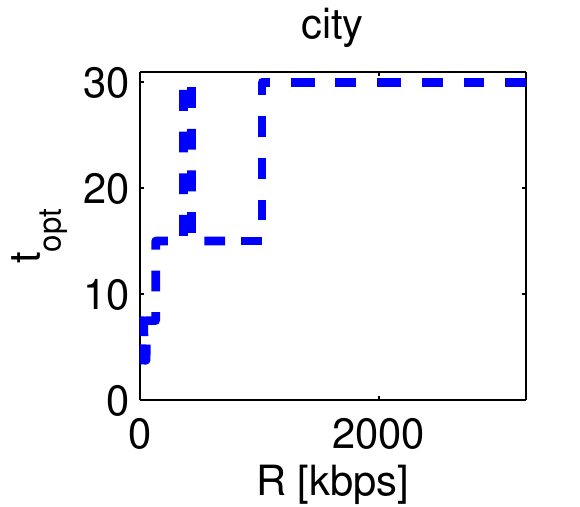}
  \includegraphics[scale=0.34]{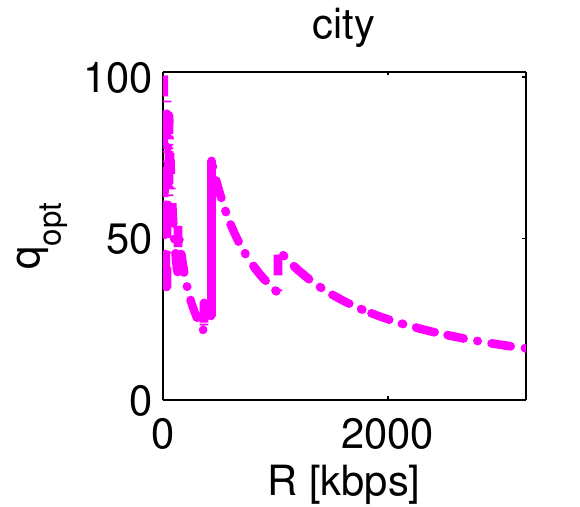}

  \includegraphics[scale=0.34]{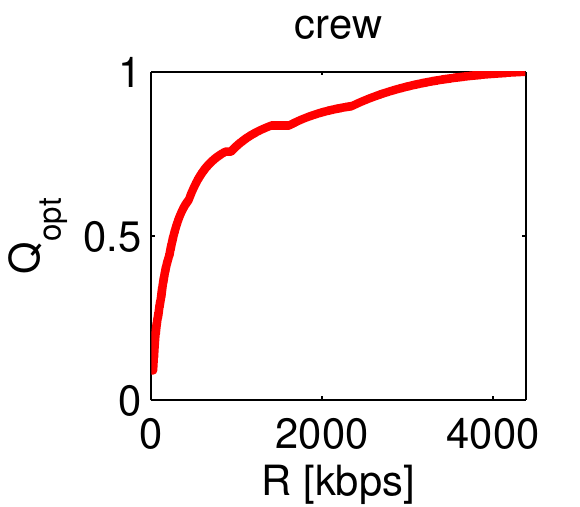}
  \includegraphics[scale=0.34]{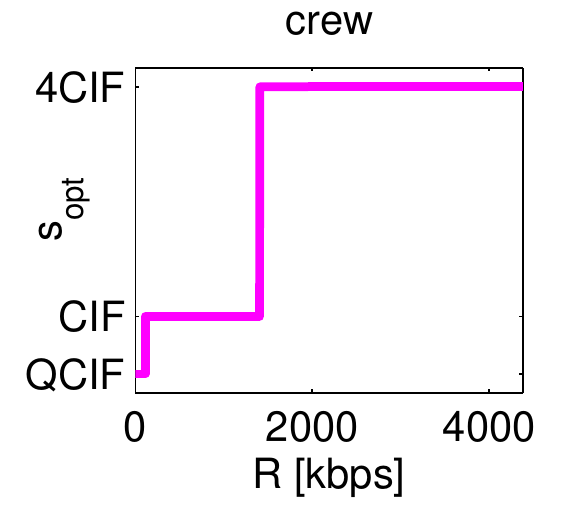}
  \includegraphics[scale=0.34]{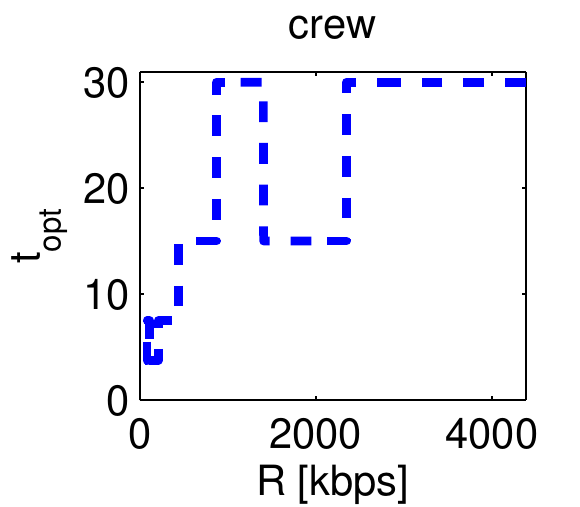}
  \includegraphics[scale=0.34]{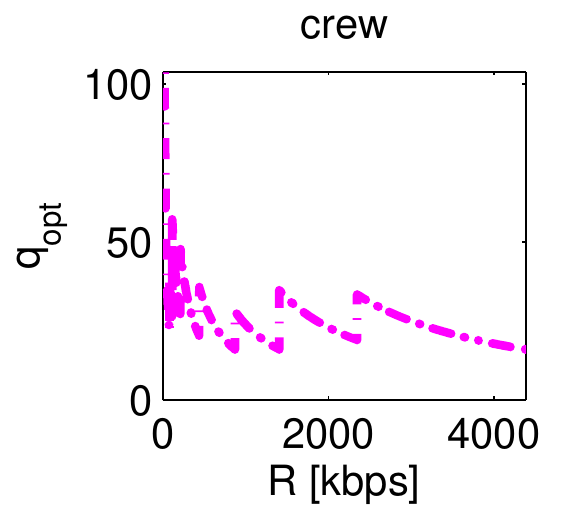}

  \includegraphics[scale=0.34]{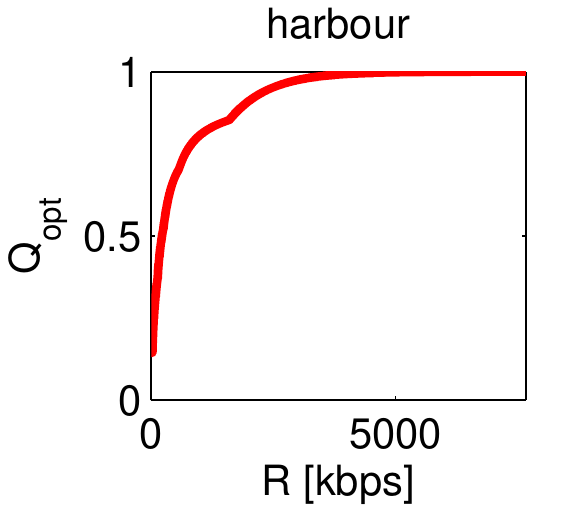}
  \includegraphics[scale=0.34]{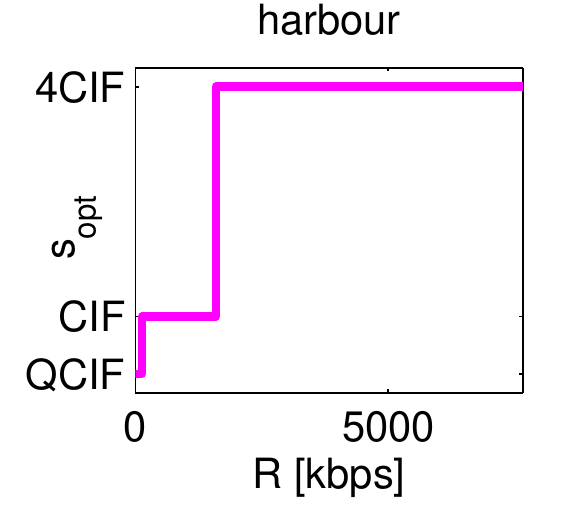}
  \includegraphics[scale=0.34]{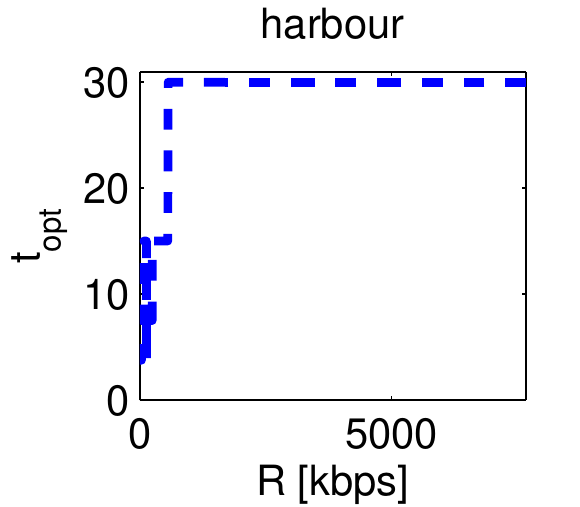}
  \includegraphics[scale=0.34]{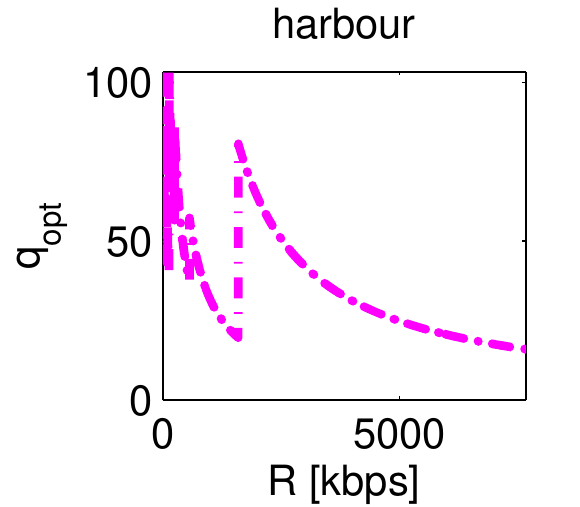}

  \includegraphics[scale=0.34]{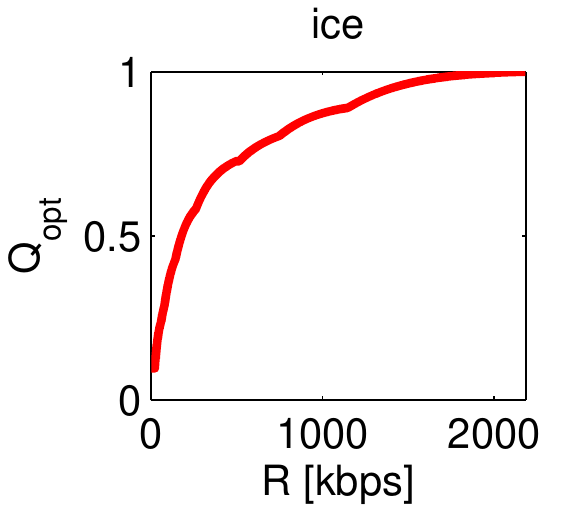}
  \includegraphics[scale=0.34]{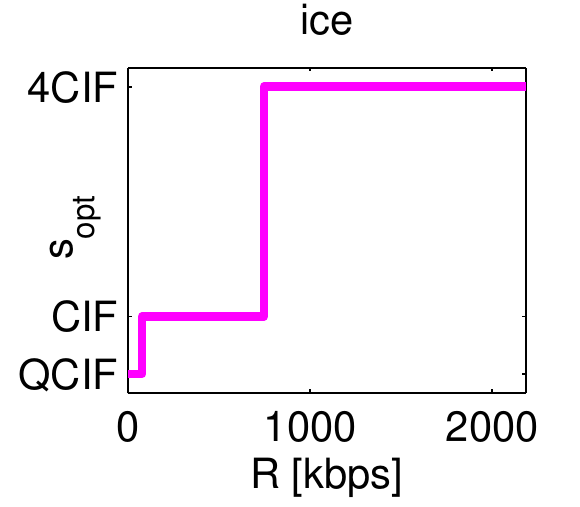}
  \includegraphics[scale=0.34]{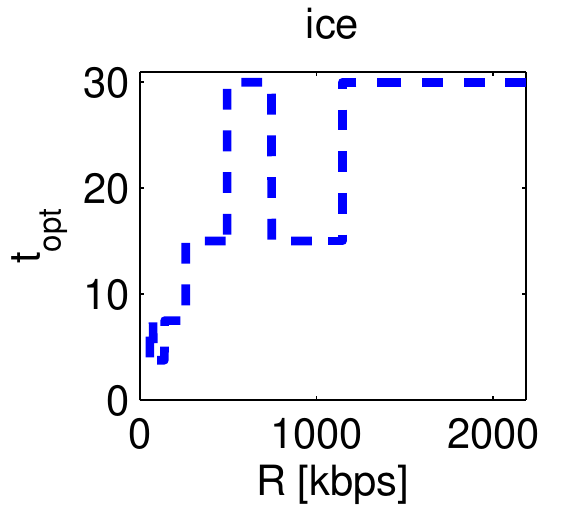}
  \includegraphics[scale=0.34]{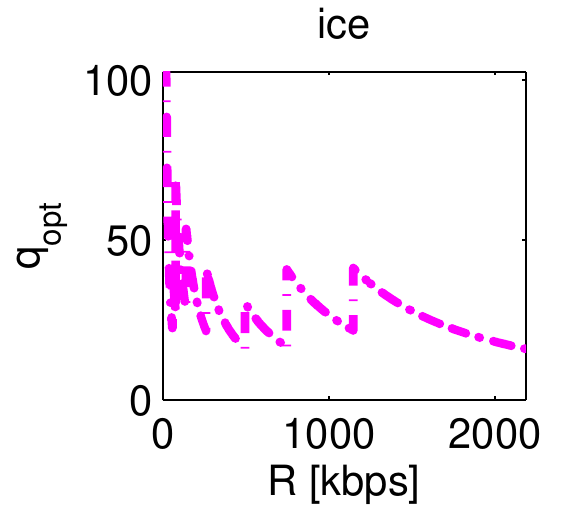}

  \includegraphics[scale=0.34]{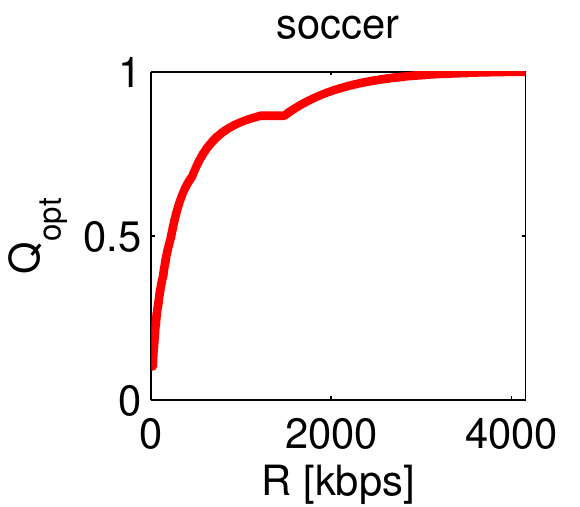}
  \includegraphics[scale=0.34]{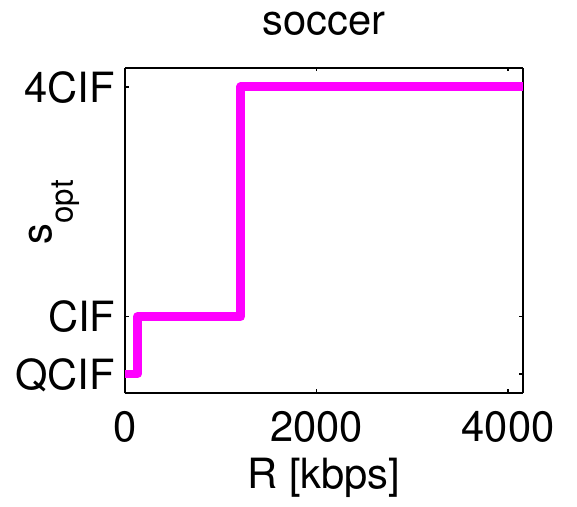}
  \includegraphics[scale=0.34]{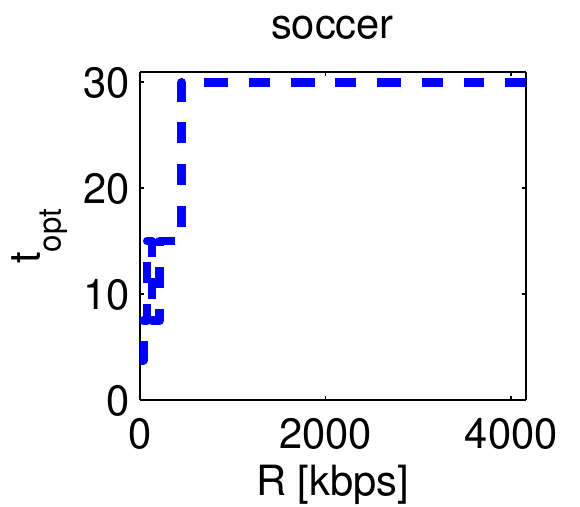}
  \includegraphics[scale=0.34]{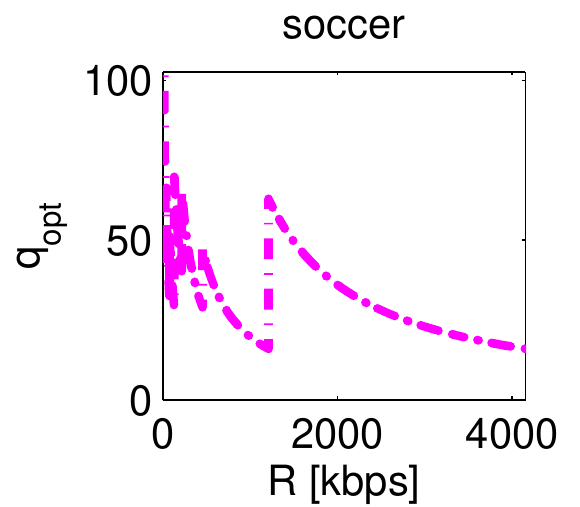}

  \caption{Optimal $Q$, $s$, $t$ and $q$ v.s. $R$ by assuming continuous $q$,
  $t\in\{3.75, 7.5,15,30\}$ Hz and $s\in\{\mbox{QCIF, CIF, 4CIF}\}$ .}
  \label{fig:RQSTAR_discrete}

\end{figure}

In the scalable video adaptation problem considered in the previous
section, we assume that all SVC layers are stored in a server (or
sender), and for a given rate, the layers corresponding to the
optimal STAR for that rate are extracted and sent. In this scenario,
the optimal STAR  corresponding to increasing rates does not need to
be monotonically increasing. As shown in
Fig.~\ref{fig:RQSTAR_discrete}, in fact, with limited choice of
feasible $s$ and $t$, the optimal STAR indeed are not monotonically
increasing.

Here, we consider a different scenario, where we would like to
preorder the SVC layers (coded using discrete sets for $s$, $t$, and
$q$) into a single layered stream, so that each additional layer
yields the maximum possible quality improvement. With such a
pre-ordered SVC stream, the server or proxy in the network can
simply keep sending additional layers, until the rate target is
reached.   The order of feasible STAR points must satisfy the
monotonicity  constraint that, as rate increases, $s$ and $t$ be
non-decreasing and $q$  be non-increasing.

We have shown previously how to employ \eqref{eq:RSTAR} and
\eqref{eq:QSTAR} to determine the optimal STAR which gives the best
quality under a rate constraint. There we have assumed either $s$,
$t$, and $q$ are all continuous, or $s$ and $t$ are discrete but $q$
is continuous. In practice,  $q$ as well as $s$ ant $t$ can only be
chosen from a finite set, so that the achievable rate is not
continuous. Although we could extend the scheme in Sec.
\ref{ssec:discrete_s_t} to allow only discrete $q$, the algorithm
will not be very efficient, as we don't know what are the achievable
rates in advance. Also, the resulting $(s,t,q)$ points may not
satisfy the monotonicity constraint. In the following, we discuss
how to take into account such practical limitations.

\begin{figure}
\centering
{\includegraphics[scale=0.45]{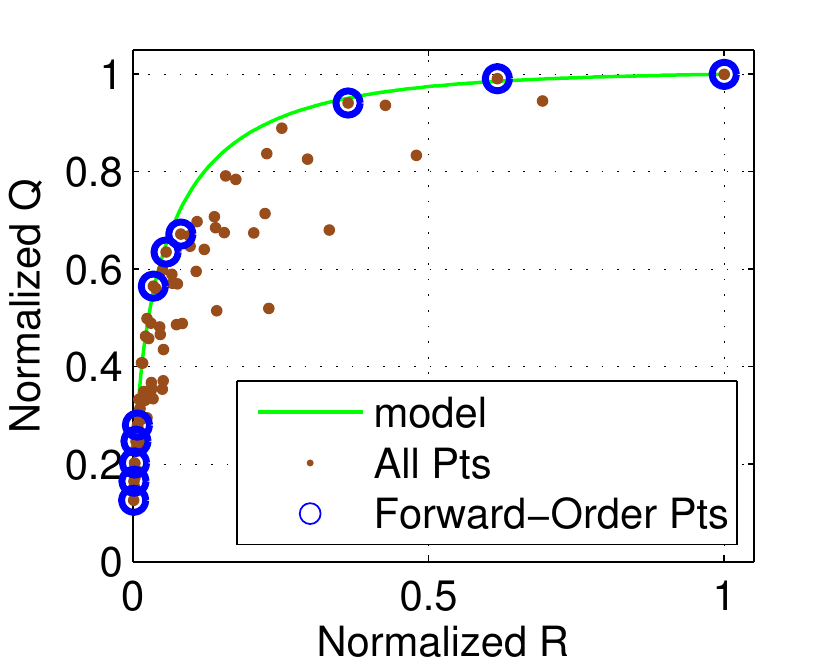}}
{\includegraphics[scale=0.45]{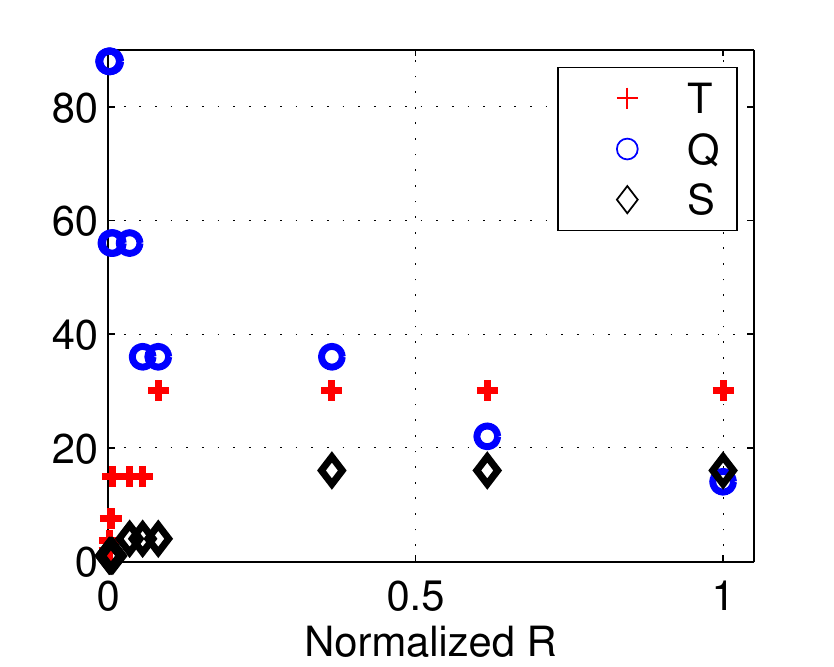}}
{\includegraphics[scale=0.45]{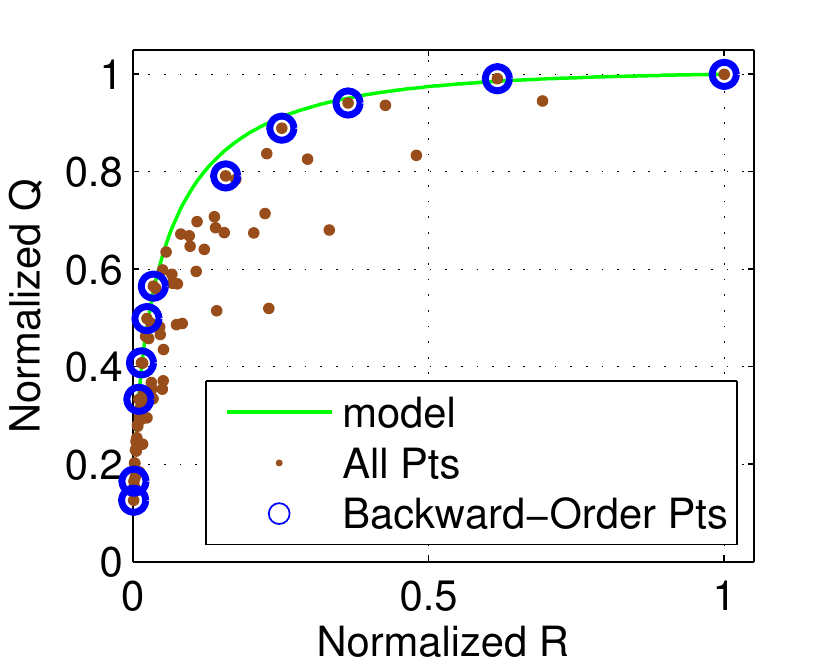}}
{\includegraphics[scale=0.45]{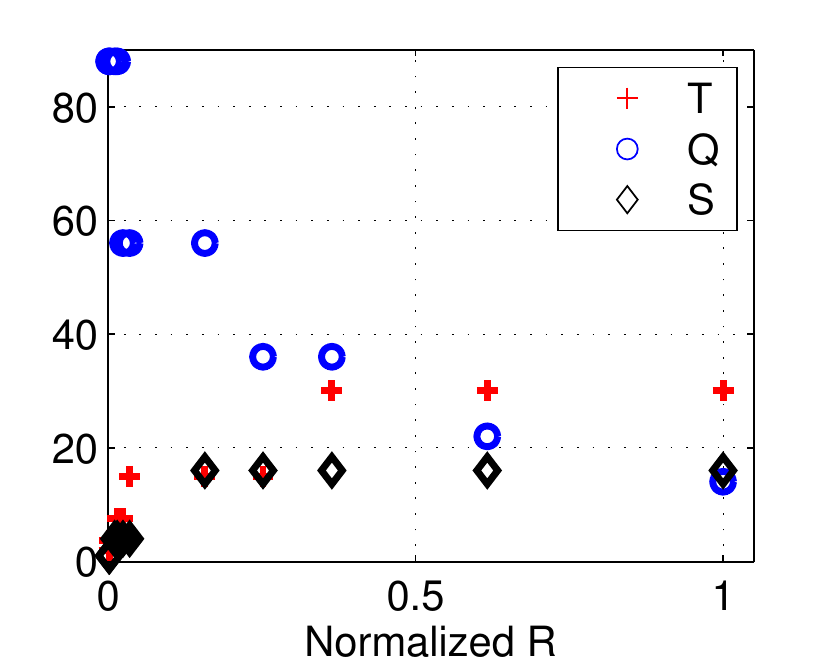}}
\caption{\label{fig:quality_opt_points} Quality optimized pre-order
points for ``city''.}

\end{figure}

Suppose there are $L$ spatial layers, $M$ temporal layers and $N$
amplitude layers, the corresponding feasible choices of FS, FR and
QS are ${\cal S} = \{s_1, s_2, ..., s_L\}$, ${\cal T} = \{t_1, t_2,
... , t_M\}$ and ${\cal Q} = \{q_1, q_2, ... , q_N\}$, respectively,
where FS and FR are increasingly ordered and QS is decreasing
ordered. We denote each combination of STAR and the associate rate
and quality as $(R_{lmn}, Q_{lmn}, s_l, t_m, q_n)$, $1\leq l\leq L,
1\leq m\leq M, 1\leq n\leq N$. A greedy algorithm ({\it forward
ordering algorithm}) works as follows: starting from the base layer,
i.e., $(R_{111}, Q_{111}, s_1, t_1, q_1)$, check the three possible
moves to next rate point, i.e., $(R_{211}, Q_{211}, s_2, t_1, q_1)$
or $(R_{121}, Q_{121}, s_1, t_2, q_1)$ or $(R_{112}, Q_{112}, s_1,
t_1, q_2)$. Move to the one with maximum quality gain over the rate
increase, i.e., $\operatorname{arg\,max}\limits_{l, m, n} \Delta Q/ \Delta R$. 
The process continues until $(s,t,q)=(s_L, t_M, q_N)$.
Fig.~\ref{fig:quality_opt_points}(a) shows the achievable
rate-quality points for ``city'' obtained using this forward
pre-order algorithm. We can see that those points follow the
continuous rate-quality model \eqref{eq:QR} very closely, indicating
that ordering SVC layers subjecting to the monotonicity  constraint
yields near-optimal rate-quality tradeoff. However, due to the
coarse granularity of feasible $s$, $t$, and $q$ in each move, it
results in a ``clustered'' rate points.
Fig.~\ref{fig:quality_opt_points}(b) shows the corresponding optimal
$(q, s, t)$ as functions of rates. It is easy to confirm that the
monoticity of $(q, s, t)$ is satisfied\footnote{We assume spatial
resolution at QCIF is 1, and normalize CIF and 4CIF as 4 and 16.}.

A problem with the forward ordering method is that the achievable
rate points tend to cluster in separate regions, leaving relatively
large ``gaps'' in the rate region $(0, R_{\max})$. To generate a
more uniformly distributed set of rates, we also design a {\it
backward ordering algorithm}. The algorithm works similar as the
forward order algorithm, except that it starts with the last points,
i.e. $(R_{LMN}, Q_{LMN}, s_L, t_M, q_N)$, and proceeds ``backward''
to the first points $(R_{111}, Q_{111}, s_1, t_1, q_1)$. We also
consider the three possible moves along spatial, temporal or
amplitude dimensions, and keep the one with minimum quality drop
over the rate drop.

\begin{figure}[htp]
  \centering
  \includegraphics[width=0.4\columnwidth]{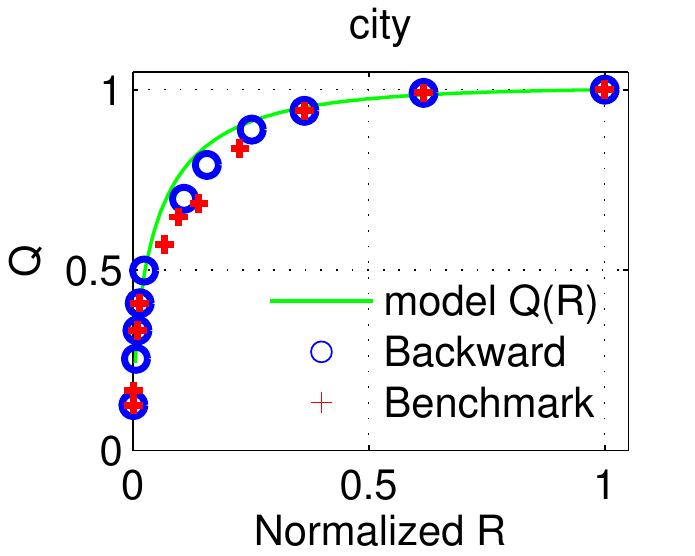}
  \includegraphics[width=0.4\columnwidth]{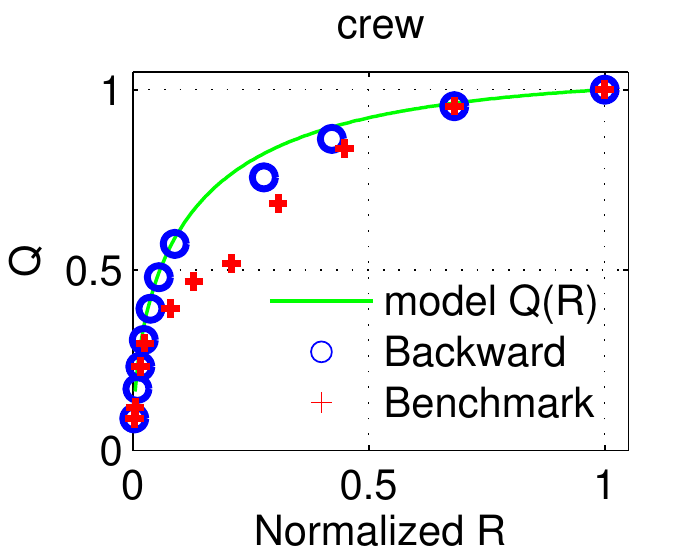}
  \includegraphics[width=0.4\columnwidth]{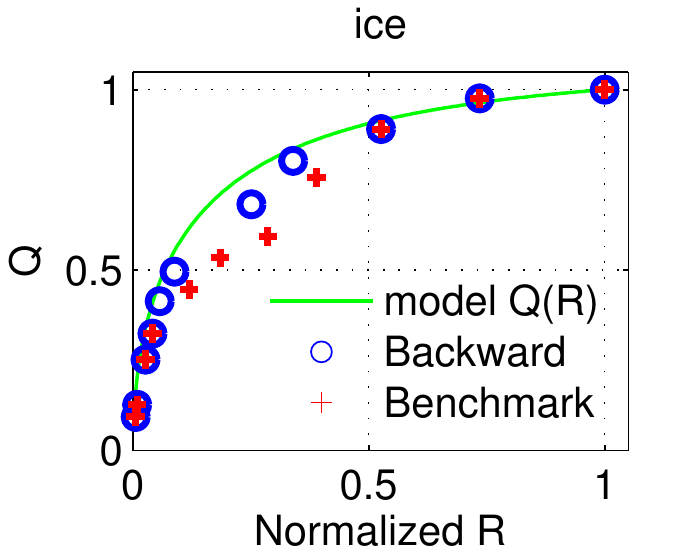}
  \includegraphics[width=0.4\columnwidth]{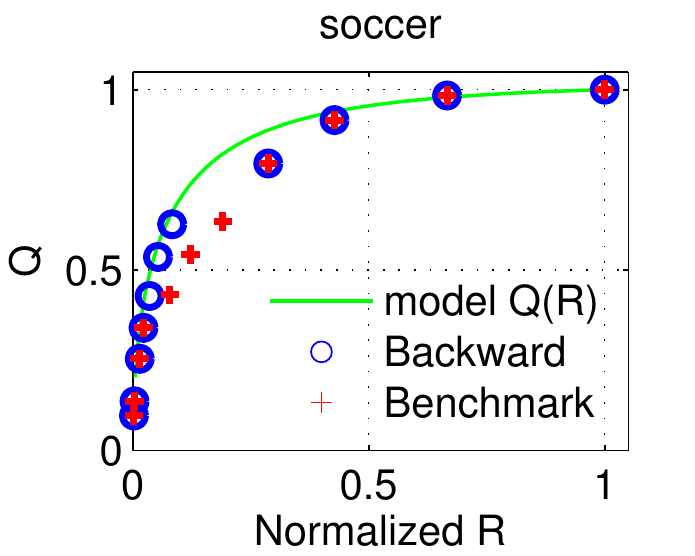}
  \caption{\label{fig:order_compare} Achievable rate-quality points by  forward and backward algorithms.}

\end{figure}

Fig.~\ref{fig:quality_opt_points}(c) and
Fig.~\ref{fig:quality_opt_points}(d) show  the  achievable
rate-quality points using the backward algorithm and the
corresponding STAR for ``city''. Clearly, the points follow the
continuous rate-quality model \eqref{eq:QR} very closely.
 Furthermore, they are more uniformly distributed over the entire rate region, compared to the points obtained by the forward-ordering algorithm show in Fig.~\ref{fig:quality_opt_points}(a).  The
performance comparison for other sequences between the backward
algorithm and the forward algorithm is shown in
Fig.~\ref{fig:order_compare}. Clearly, both algorithms return
near-optimal pre-order points. However, the backward algorithm
provides more uniformly distributed rate points.

\section{Discussion and Conclusion} \label{sec:conclusion}
In this paper, we propose an analytical rate model considering the
impact of the spatial, temporal and amplitude resolutions (STAR). We
have found the impact of spatial, temporal and amplitude resolution
on video bit rate is actually separable. Hence the rate model
consider STAR combinations is the product of the separate functions
of frame rate, frame size and quantization stepsize. Our proposed
analytical rate model is generally applicable to all coding
scenarios, including both scalable and single layer video coding,
using hierarchical B or IPPP for temporal prediction, with or
without QP cascading, etc. But the model parameters differ depending
on the encoding scenarios. We have also verified that our model is
accurate for other video contents (e.g., test videos from JCT-VC)
and resolutions (e.g., 720p, WVGA, etc), which are not included here
because of the space limitation.

Experimental results show that model parameters are highly content
dependent. We also propose a method for predicting the model
parameters using weighted sum of some content features that can be
computed from original video sequences. We have found that it is
sufficient to provide accurate bit rate estimation with model
parameters predicted by three content features. We also notice that
the best feature set is the same for all test cases, but the
predictor matrix {\bf H} depends on the encoder setting.

Finally, we show how to use the proposed rate model and a
corresponding quality model, both as functions of $q$, $s$, $t$,  to
determine the  STAR that maximizes the perceptual quality for a
given rate constraint. The solution is applicable to both encoder
rate control, and scalable video adaptation.  We further show that
the rate and quality models  can be combined to order the layers in
a scalable stream in a rate-quality optimized way.

\bibliographystyle{IEEEtran}
\bibliography{SmartAdaptation}

\end{document}